\documentclass[12pt, oneside]{article}   	
\usepackage[margin=1in]{geometry}
\usepackage[parfill]{parskip}    			
\usepackage{graphicx}				
\usepackage{amsthm,amssymb,amsmath}
\usepackage{mathtools}
\usepackage{comment}
\usepackage{enumerate,enumitem} 
\usepackage{bbm}
\usepackage{tabularx}
\usepackage[english]{babel}
\usepackage{amsthm}
\usepackage{multirow}
\usepackage[table,xcdraw]{xcolor}
\usepackage{subcaption}
\usepackage[labelformat=parens,labelsep=quad, skip=3pt]{caption}
\usepackage[normalem]{ulem}
\useunder{\uline}{\ul}{}

\newtheorem{proposition}{Proposition}

\usepackage{tikz}
\usepackage{natbib}
\usepackage{subcaption}
\usetikzlibrary{matrix,positioning}
\tikzset{bullet/.style={circle,fill,inner sep=2pt}}
\usetikzlibrary{arrows}

\begin{document}
\begingroup
\centering
{\LARGE  Marked Cox Models for IBNR Claims Count: Continuous and Discretized Approaches with Dirichlet-Driven Reporting Delays\\[1.5em]
\large Hassan Abdelrahman*{$^{1}$}, Andrei Badescu{$^{1}$}, Radu Craiu{$^{1}$}, Sheldon Lin{$^{1}$}}\\[1em]
$^1$Department of Statistical Sciences, University of Toronto \\[1em]
\centerline{{*hassan.abdelrahman@mail.utoronto.ca}}
\endgroup
\begin{abstract}
Accurate loss reserving is crucial in Property and Casualty (P\&C) insurance for financial stability, regulatory compliance, and effective risk management. We propose a novel micro-level Cox model based on hidden Markov models (HMMs). Initially formulated as a continuous-time model, it addresses the complexity of incorporating temporal dependencies and policyholder risk attributes. However, the continuous-time model faces significant challenges in maximizing the likelihood and fitting right-truncated reporting delays. To overcome these issues, we introduce two discrete-time versions: one incorporating unsystematic randomness in reporting delays through a Dirichlet distribution and one without.

We provide the EM algorithm for parameter estimation for all three models and apply them to an auto-insurance dataset to estimate IBNR claim counts. Our results show that while all models perform well, the discrete-time versions demonstrate superior performance by jointly modeling delay and frequency, with the Dirichlet-based model capturing additional variability in reporting delays. This approach enhances the accuracy and reliability of IBNR reserving, offering a flexible framework adaptable to different levels of granularity within an insurance portfolio.
\\
\\
\noindent{\textbf{Keywords:} Dirichlet distribution, EM algorithm, hidden Markov model, IBNR claims, marked Cox model, micro-level reserving}
\end{abstract}

\section{Introduction}

In Property and Casualty (P\&C) insurance, accurate loss reserving is essential for insurers to guarantee financial stability, regulatory compliance, and effective risk management (\cite{bjarnason2014insurance}). Loss reserving involves estimating the funds needed to cover future payments for incurred claims. Two key components of loss reserves are the Reported But Not Settled (RBNS) reserve, accounting for claims reported but not yet settled, and the Incurred But Not Reported (IBNR) reserve, accounting for incurred claims that are yet to be reported. While both components are essential, this paper focuses on IBNR reserving. Unlike RBNS claims, the number of IBNR claims is unknown to the insurer at the time of reserve valuation, necessitating sophisticated modeling techniques to accurately estimate this reserve. The complexity of IBNR reserving lies in forecasting the frequency and severity of unreported claims, making it a challenging yet crucial aspect of the insurance reserving process.

The majority of the literature on modeling insurance claims and reserving is based on aggregated claims data, such as the chain ladder method (\cite{mack1993distribution,mack1999standard}) and the Bornhuetter-Ferguson method (\cite{bornhuetter1972actuary}). These ``macro-level'' models accumulate historical data on the development of claims over time in a two dimensional table, referred to as run-off triangle, by aggregating payments by occurrence and development year. Stochastic models were subsequently introduced to account for the variability in reserves derived from these models; see \cite{taylor2012loss}, \cite{england2002stochastic}, and \cite{wuthrich2008stochastic} for a comprehensive overview of stochastic reserving.  Despite their simplicity, these models come with several disadvantages, including the loss of valuable insights into the characteristics of individual claims, high parameter uncertainty due to the limited number of observations in aggregated data leading to reduced predictive power, and potential bias under certain conditions (\cite{crevecoeur2022hierarchical}). To resolve these shortcomings, ``micro-level" models have been proposed with the aim to use individual-level data to depict the development of individual claims. 

\cite{arjas1989claims} and \cite{norberg1993prediction,norberg1999prediction} are the first to propose a stochastic loss reserving framework at individual claim level. They proposed a marked non-homogeneous Poisson process (NHPP) model and a general mathematical framework for the development of individual claims. \cite{haastrup1996claims} then implemented the NHPP model using non-parametric Bayesian statistics. The subject of micro-level reserving remained unopened before emerging back in recent years. \cite{antonio2014micro} presented a case study in which they applied Norberg's model to an auto-insurance dataset. Data-driven comparisons between macro-level and micro-level reserving have shown the superiority of the micro-level models; see \cite{charpentier2016macro} and \cite{huang2015stochastic,huang2016asymptotic}. 

Since then, various research streams have emerged in the micro-level estimation of IBNR reserves. One stream employs machine learning and neural network techniques for IBNR claim estimation. As outlined by \cite{bucher2023micro}, this approach utilizes Frequency-Severity or Chain Ladder-based methods to estimate IBNR reserves over discrete time intervals, as demonstrated by studies such as \cite{wuthrich2018machine,wuthrich2018neural}, \cite{baudry2019machine}, \cite{de2019claim}, \cite{delong2022collective}, and \cite{bucher2023micro}. Another stream focuses on modeling the claim arrival process, where the NHPP model has traditionally played a prevalent role. However, considerations for enhancing this aspect of loss reserving modeling have been highlighted, as discussed by \cite{badescu2016marked}. NHPP, assuming independence among claim numbers from different periods, fails to capture dependence among individual claim arrivals due to environmental variations affecting the entire portfolio (\cite{grandell2012aspects}) and contradicts the calendar year effect seen in the run-off triangle (\cite{holmberg1994correlation} and \cite{shi2012bayesian}).

Addressing these issues involves incorporating a temporal dependence structure into the claim arrival process model, achievable by transitioning to a Cox process. Notable research in this direction includes \cite{avanzi2016micro}, who utilized the shot noise Cox process to model claim arrival, employing a reversible jump Markov Chain Monte Carlo method for estimation. Conversely, \cite{badescu2016marked} employed a hidden Markov model (HMM) with an Erlang state-dependent distribution to model the claim arrival process for the entire portfolio. Their work showcased that the discretely observed claim arrival process, along with associated reported claims and IBNR processes, follows a Pascal-HMM. The authors detailed the estimation algorithm for their model and its application in estimating IBNR claim count in \cite{badescu2019marked}. It is crucial to note that while both studies incorporated the Cox model, they focused on modeling the claim arrival process for the entire portfolio, overlooking policyholder information that is likely to enhance the accuracy of the different components within the model.

To this end, we introduce a novel micro-level Cox model that offers a level of flexibility allowing actuaries to customize the analysis to various levels of granularity. Specifically, our model accommodates modeling at the policyholder level, lines of business level, or the entirety of the portfolio, providing a comprehensive framework for analyzing claim arrival patterns. In the same spirit as \cite{badescu2016marked}, we adopt an HMM as the foundation of our Cox model. In our framework, we assume a shared hidden Markov process governing the claim arrival processes across different granular units within the portfolio. From a policyholder-level modeling perspective, this HMM serves as an environmental dynamic that simultaneously influences all policyholders. Given a state of the HMM, our model assumes that the claim arrival intensity for a policyholder is a constant, which depends on the risk attributes of the policyholder via a regression function. Thus, the claim arrival intensity for a policyholder is influenced by both external factors affecting all policyholders and their own risk attributes. The temporal dependence and environmental variation, often not incorporated in most reserving models, are captured through the common hidden Markov process. We show that within our framework, the discretely observed claim arrival process, the discretely observed reported claim process, and the discretely observed IBNR claim process for each policyholder belong to the family of Poisson-HMMs, where the state-dependent distribution depends on the policyholder's risk attributes. 

We initially propose our model in continuous time. In general, fitting continuous marked count process models to reported claims data (whether they are marked Poisson or marked Cox processes) presents significant challenges. One of the main challenges is maximizing the likelihood of observed data. Maximizing the likelihood of the observed data entails addressing two distinct components: the likelihood for the truncated reporting delay and the likelihood for the claim frequency model of reported claims, which in turn depends on the reporting delay. Thus, maximizing the likelihood poses a formidable task due to the interdependence of the two components. In recent micro-level reserving literature, only \cite{wahl2019explicit} have attempted to maximize the likelihood at once for a specific class of micro-level reserving models. Traditionally, researchers have adopted a two-step maximization approach to maximize the likelihood of the observed data, initially fitting the right-truncated reporting delay and subsequently estimating the frequency model using the reporting-delay components in the frequency model as constants (which are estimated from the first step). However, this approach rests on the assumption of independence between claim frequency and reporting delay, which is often invalid; an increase in reported claims may indicate either a surge in actual claims occurrence, accelerated reporting, or a combination of both, challenging the independence assumption.

We address such challenge within our marked Cox model framework by integrating reporting delay and claim frequency modeling cohesively, transitioning from a continuous-time framework to a discrete-time framework. Specifically, we aggregate continuous-time events to a coarser time scale, which is more common in practice. This coarser time scale (or period) can be days, weeks, months, etc. We assume a hierarchical structure: if $N_{i,t}$ denotes the number of claims that occurred in period $t$ for policyholder $i$, and $Z_{i,t,d}$ denotes the number of these claims reported $d$ periods later, then we assume that $Z_{i,t,d} \mid N_{i,t}$ follows a multinomial distribution. We simultaneously model both the delay and occurrence components, capturing the intricate interplay between claim occurrence and reporting behavior.

Moreover, existing literature often assumes either static or time-dependent covariates to model reporting delay variations. However, practical scenarios frequently involve unexplainable changes in delay structures that would not be captured through conventional covariates. As previously demonstrated by \cite{swamy1970efficient} and \cite{csenturk2005covariate}, accounting for such unexplainable changes requires us to assume that the delay probabilities are random. To address this, we extend the discrete-time model to account for the likely occurrence of unobserved confounding covariates by assuming that the probabilities of the multinomial distribution follow a Dirichlet distribution.

We outline the EM algorithm needed to obtain the maximum likelihood estimates for the three different models: the continuous-time model assuming two-step maximization of the observed data likelihood, and the discrete-time models, both with and without the Dirichlet assumption for the reporting delay probability vector. We also fit the three models to an auto-insurance dataset from a major European insurance company, which exhibits significant randomness in the occurrence and delay structure. We show that despite losing information by switching to the discrete-time framework, the discrete-time models perform better than the continuous-time model because they account for the joint modeling of occurrence and reporting. Moreover, we demonstrate that given the randomness exhibited in our data, adding the Dirichlet assumption provides more realistic interval estimates for the IBNR claim count.

This work contributes to the literature by:
\begin{itemize}[leftmargin=*]
    \item Extending the Cox model of \cite{badescu2016marked,badescu2019marked} to various levels of granularity (e.g., policyholder level, lines of business level). This is specifically important if the distribution of granular units in our portfolio changes over time and these granular units possess different characteristics in terms of claim occurrence, reporting delay, and claim severity. Ignoring these differences would lead to biased reserve estimates.
    
    \item Building upon the work of \cite{verbelen2022modeling} on modeling event occurrence subject to reporting delay via an EM algorithm, where they assume a Poisson process for claim occurrence. We extend their analysis by allowing the occurrence process to follow a Poisson-HMM, as well as conducting our analysis from a micro-level perspective.

    \item Incorporating unsystematic randomness in the reporting delay probability vector by assuming that it follows a Dirichlet distribution, and outlining the EM algorithm needed to fit the model. We show that this is also an extension to the work of \cite{verbelen2022modeling} who do not assume randomness of the reporting delay. 
\end{itemize}

The paper is structured as follows: In Section \ref{sec:framework}, we outline our modeling framework and discuss its interpretability. The properties of the model and the associated individual-level reported claim process and IBNR process are presented in Section \ref{sec:prop}. In Section \ref{sec:likelihood}, we outline the discrete-time version of the model and show the likelihood for the three different models under consideration. In Section \ref{sec:est}, we provide the EM algorithm needed to estimate the parameters of the three different models. Section \ref{sec:ibnr} details the application of the model in predicting the IBNR claims count and reserve, while we apply the models to real-life data in Section \ref{sec:data} to assess their performance in estimating the IBNR claim count. We conclude in Section \ref{sec:conc}.

\section{Modeling Framework} \label{sec:framework}
We present our proposed model for modeling the claim arrival process at the policyholder level. In line with the notation established by \cite{badescu2016marked}, we describe the development of the $j$th claim for the $i$th policy $(i=1,\dots,m)$ through three random variables $(T_{ij},U_{ij},\boldsymbol{Z}_{ij})$, where $T_{ij}$ denotes the occurrence time of the claim, $U_{ij}$ represents its reporting delay, and $\boldsymbol{Z}_{ij}$ characterizes the development process following claim reporting. Chronologically ordered, $\{(T_{ij},U_{ij},\boldsymbol{Z}_{ij}), j=1,2,\dots\}$ constitutes the claim history process for policy $i$, whose risk attributes are denoted by $\boldsymbol{x}_{i}$. Additionally, the total claim count process for the $i$th policy is defined as $N_i(t)=\sum_{j=1}^{\infty}\mathbbm{1}\{T_{ij}<t\}$, where $\mathbbm{1}\{.\}$ is the indicator function. 

In our modeling approach, we represent $\{N_i(t), t\ge 0 \}$ as a marked Cox process with two components:

\begin{enumerate}
    \item The marks $\{(U_{i1},\boldsymbol{Z}_{i1}),(U_{i2},\boldsymbol{Z}_{i2}),\dots\}$, and
    \item The stochastic intensity function $\Lambda_i(t)$.
\end{enumerate}

The assumption is made that the marks are independent, with a common density function given by $f_{U,\boldsymbol{Z}\mid t,\boldsymbol{x}_i}(u,\boldsymbol{z})=f_{U\mid t,\boldsymbol{x}_i}(u)f_{\boldsymbol{Z}\mid u,t,\boldsymbol{x}_i}(\boldsymbol{z})$. Additionally, the stochastic intensity function $\Lambda_i(t)$ is modeled as a piecewise stochastic process, where $\Lambda_i(t) = e_{i,l}\Lambda_l(t;\boldsymbol{x}_i)$ for $d_{l-1} \le t < d_l$, with $l=1,2,\dots$ and $d_0=0$. The time points $d_l$, $l=1,2,\dots$, are predetermined, and $t=d_0=0$ marks the beginning of the observation window for the entire portfolio. Moreover, $e_{i,l}\in[0,1]$ denotes the exposure of the $i$th policy in the interval $d_{l-1} \le t < d_l$, given by $e_{i,l}=\frac{\#\text{ days for which the contract is in force in }d_{l-1} \le t < d_l}{\#\text{ days in }d_{l-1} \le t < d_l}$. Notably, $e_{i,l}$ can be zero, indicating the case when the $i$th policy's contract is not active during $d_{l-1} \le t < d_l$.

To describe $\{\Lambda_1(t;\boldsymbol{x}_i), \Lambda_2(t;\boldsymbol{x}_i),\dots\}$, we propose a structure involving two components:

\begin{enumerate}
    \item A hidden parameter process $\{C_1,C_2,\dots\}$, which constitutes a time-homogeneous Markov chain with a finite state space $\{1,\dots,g\}$. Denoting its initial distribution and transition probability matrix as $\boldsymbol{\pi}_1$ and $\boldsymbol{\Gamma}=\{\gamma_{jk}\}$, respectively, where $\gamma_{jk}=P(C_l=k\mid C_{l-1}=j)$, we assume the Markov chain is irreducible, aperiodic, and has all states being positive recurrent, leading to a unique limiting distribution.
    \item A state-dependent process $\{\Lambda_1(t;\boldsymbol{x}_i), \Lambda_2(t;\boldsymbol{x}_i),\dots\}$, which depends on the current state $C_l$. Given $C_l=j$, we assume that $\Lambda_l(t;\boldsymbol{x}_i)$ is constant such that $$\Lambda_l(t;\boldsymbol{x}_i)\mid (C_l=j) = \lambda^{(j)}(\boldsymbol{x}_i),$$ where $\lambda^{(j)}:\mathcal{X}\to \mathbb{R}^+$ is a regression function, and $\mathcal{X}$ is the covariate space.
\end{enumerate}

Specifically, each policyholder's risk attributes $\boldsymbol{x}_i$ are incorporated into the model through the regression function $\lambda^{(j)}(\boldsymbol{x}_i)$. These risk attributes could include factors such as age, location, coverage type, and any other relevant information about the policyholder that may affect claim occurrence rates. By including these covariates, we tailor the intensity function $\Lambda_l(t;\boldsymbol{x}_i)$ to reflect the unique risk profile of each policyholder.

In summary, the intensity for each policy $i$ remains constant during each period $l$, contingent on the realization of hidden parameter $C_l$, and is determined by the regression function based on policyholder's risk attributes. The hidden states represent environmental variation (e.g. weather or seasonal effect), affecting all policies in the portfolio, with policies assumed independent given the environmental variation. It is important to note that the state of the $l$th period is identical for all policies $i$. Consequently, we have that, for each policy $i$ $(i=1,\dots,m)$, 
\[ \Lambda_l(t;\boldsymbol{x}_i) = \begin{cases} 
      \lambda^{(1)}(\boldsymbol{x}_i), & \text{with prob. = }\pi_{l1}\\
      \vdots & \vdots \\
      \lambda^{(g)}(\boldsymbol{x}_i), & \text{with prob. = }\pi_{lg}
   \end{cases}, \quad t \in [d_{l-1},d_l).
\]

\textbf{Remarks}
\begin{itemize}[leftmargin=*]
    \item It is important to note that our proposed Cox model framework generalizes the work of \cite{badescu2016marked}, albeit with different modeling assumptions. While \cite{badescu2016marked} models the claim arrival process for the entire portfolio, our approach takes a more granular stance by considering the claim arrival process for each individual policyholder (or other granular units), incorporating their respective risk covariates $\boldsymbol{x}_i$. Thus, our model accounts not only for seasonality or environmental variations but also explicitly considers the risk characteristics of individuals, allowing the number of claims to depend on both external factors and the unique risk profiles within the portfolio. 
    
    Moreover, in the work of \cite{badescu2016marked}, the intensity function for a time period $[d_l,d_{l+1})$ is assumed to follow an Erlang distribution. While this assumption might be appropriate for a macro-level perspective, it is less suitable for our micro-level approach, where we employ a regression function to model the intensity. Assuming the intensity is constant and dependent on covariates simplifies our setup and aligns more closely with existing literature. In the next section, we show that the distribution of the number of claims in a given period under our model will be a mixture of Poisson distributions, which a good fit for the over-dispersed claim frequency data. Assuming an Erlang distribution for the intensity would result in a mixture of Pascal distributions for the number of claims, which is not expected to significantly enhance model performance. Additionally, with appropriate choices of predetermined time points, the intensity can be safely assumed to be constant.

    \item A special case within the above framework arises when the practitioner aims to model the claim arrival process across different lines of business. In this scenario, the index $i$ corresponds to the $i$th line of business. If the regression function $\lambda^{(j)}(.)$ is a constant-regression function, our model aligns with the contemporaneously conditionally independent multivariate HMM; see \cite{macdonald2016hidden}. Importantly, our framework permits the inclusion of portfolio-level information through covariates in the regression model (e.g., the average age of policyholders in the line of business). Many papers argue for the necessity of accounting for the dependence between different lines of business in reserving practices; the HMM captures this dependence via hidden states that influence the claim arrival process across all lines of business. Conversely, if the practitioner models the claim arrival process for a single line of business, they may choose $i$ to correspond to the $i$th claim type, thereby modeling claim arrivals for different claim types. This distinction is crucial because even within the same line of business, the distribution of the marks may vary across different claim types. 
    
    \item An essential assumption in our model is that, given a realization of the hidden state during $[d_{l-1},d_l)$, the intensity function is considered constant - with respect to an individual's risk attributes - within that period. While data-driven choices for $d_l$ could ideally capture variations in intensity over time, such an approach might overcomplicate the model and lead to overfitting issues. Setting $d_l$ to weekly, biweekly, or monthly intervals provides a reasonable balance between granularity and simplicity.
    
    \item In our proposed framework, risk attributes $\boldsymbol{x}_i$ are not explicitly time-indexed, aligning with the common practice where policyholders' risk characteristics are assumed to remain constant over time. It's crucial to recognize that, in practice, insurance risk attributes can dynamically change. For example, in auto insurance, age is often a risk attribute that changes upon policy renewal. However, this standard assumption should have no effect on the performance of our model.
    
    \item Similar to the model in \cite{badescu2016marked}, it is evident that both the micro-level mixed Poisson process and the Ammeter process (\cite{ammeter1948generalization}) are special cases of the proposed model.
 
\end{itemize}

\section{Properties of the Model} \label{sec:prop}
In this section, we outline the immediate properties of our model, focusing on its application from the policy-holder level. At a given reserve valuation date $\tau$, the complete claim arrival process is not fully observed; only reported claims are observable. Denoting the reported claim process for the $i$th policy with respect to $\tau$ as $\{N_i^{\text{r}}(t); 0\le t\le \tau\}$, it is defined as $N_i^{\text{r}}(t) = \sum_{j=1}^{\infty}\mathbbm{1}\{T_{ij} < t; T_{ij}+U_{ij}\le \tau\}$, where $T_{ij}$ represents the claim occurrence time and $U_{ij}$ denotes the reporting delay for the $j$th claim of the $i$th policyholder. Claims not reported by $\tau$ (i.e., $T_{ij}+U_{ij}>\tau$) contribute to the IBNR claim process, denoted as $\{N_i^{\text{ibnr}}(t); 0 \le t \le \tau\}$, and defined as $N_i^{\text{ibnr}}(t) = \sum_{j=1}^{\infty}\mathbbm{1}\{T_{ij} < t; T_{ij}+U_{ij} > \tau\}$. Although both processes should be indexed by $\tau$, we omit it for simplicity.

It's notable that both the reported claim process and the IBNR claim process remain marked Cox processes, retaining easily convertible stochastic intensity functions and mark densities. We revisit a proposition proven in Theorem 3.1 of \cite{badescu2016marked} to demonstrate this.

\begin{proposition} \label{prop1}
Assume that the claim arrival process for the $i$th policyholder, $\{N_i(t);t\ge0\}$, is a marked Cox process with stochastic intensity function $\Lambda_i(t)$ and independent marks $\{(U_{i1},\boldsymbol{Z}_{i1}),$ $(U_{i2},\boldsymbol{Z}_{i2}),\dots\}$ with common density function $f_{U,\boldsymbol{Z}\mid t,\boldsymbol{x}_i}(u,\boldsymbol{z})=f_{U\mid t,\boldsymbol{x}_i}(u)f_{\boldsymbol{Z}\mid U,t,\boldsymbol{x}_i}(\boldsymbol{z})$. Then for a given valuation date $\tau$, its associated reported claim process $\{N_i^{\text{r}}(t); 0 \le t \le \tau\}$ and IBNR claim process $\{N_i^{\text{ibnr}}(t); 0 \le t \le \tau\}$ are also marked Cox processes. Their adjusted stochastic intensity functions are $\Lambda_i^{\text{r}}(t)=\Lambda_i(t)F_{U}(\tau-t)\mathbbm{1}\{0\le t\le \tau\}$ and $\Lambda_i^{\text{ibnr}}(t)=\Lambda_i(t)(1-F_{U}(\tau-t))\mathbbm{1}\{0\le t\le \tau\}$, respectively, and their independent marks follow adjusted position-dependent mark distributions $f_{U,\boldsymbol{Z}\mid t,\boldsymbol{x}_i}^\text{r}(u,\boldsymbol{z})=\frac{f_{U\mid t,\boldsymbol{x}_i}(u)\mathbbm{1}\{u>\tau-t\}}{F_{U\mid t,\boldsymbol{x}_i}(\tau-t)}f_{\boldsymbol{Z}\mid U,t,\boldsymbol{x}_i}(\boldsymbol{z})$ and $f_{U,\boldsymbol{Z}\mid t,\boldsymbol{x}_i}^{\text{ibnr}}(u,\boldsymbol{z})=\frac{f_{U\mid t,\boldsymbol{x}_i}(u)\mathbbm{1}\{u\le\tau-t\}}{1-F_{U\mid t,\boldsymbol{x}_i}(\tau-t)}f_{\boldsymbol{Z}\mid U,t,\boldsymbol{x}_i}(\boldsymbol{z})$, respectively, where $F_{U\mid t,\boldsymbol{x}_i}$ is the distribution function for the reporting delay $U$ given a claim occurrence time $t$ and risk attributes $\boldsymbol{x}_i$. 
\hfill $\square$
\end{proposition}

Recall that we model the stochastic intensity function $\Lambda_i(t)$ of the claim arrival process $\{N_i(t;\boldsymbol{x}_i);t\ge 0\}$ as a piecewise stochastic process, where $\Lambda_i(t) = e_{i,l}\Lambda_l(t;\boldsymbol{x}_i)$ for $d_{l-1} \le t < d_l$, with $l=1,2,\dots$ and $d_0=0$, and the time points $d_l$, $l=1,2,\dots$, are predetermined. These predetermined time points constitute the time units which make up the time series on which the HMM is defined. We define $N_{i,l} := N_{i}(d_l) - N_i(d_{l-1})$ as the number of claims occurring during $[d_{l-1},d_l)$ for the $i$th policyholder ($i=1,\dots,m$). Hence, $\{N_{i,1},N_{i,2},\dots\}$ represents the discrete observations of the claim arrival process at these time points $d_l$ ($l=1,2,\dots$). The corresponding discrete observations of the reported claim process and the IBNR claim process with respect to a valuation date $\tau$ are denoted by $\{N_{i,1}^{\text{r}},N_{i,2}^{\text{r}},\dots\}$ and $\{N_{i,1}^{\text{ibnr}},N_{i,2}^{\text{ibnr}},\dots\}$, respectively. Notably, only $\{N_{i,1}^{\text{r}},N_{i,2}^{\text{r}},\dots\}$ is observed at time $\tau$. It is straight-forward to see that these three discretely observed processes follow Poisson-HMM. 

\begin{proposition} \label{prop2}
For the proposed Cox model for the claim arrival process of policyholder $i$, $N_i(t;\boldsymbol{x}_i)$, the discretely observed claim arrival process, $\{N_{i,1},N_{i,2},\dots \}$, the discretely observed reported claim process, $\{N_{i,1}^{\text{r}},\dots,N_{i,k}^{\text{r}} \}$, and the discretely observed IBNR claim process, $\{N_{i,1}^{\text{ibnr}},\dots,N_{i,k}^{\text{ibnr}}\}$, all fall under the class of Poisson-HMMs. They share the same hidden parameter process $\{C_1,C_2,\dots\}$ with $N(t;\boldsymbol{x}_i)$. Furthermore, their state-dependent distributions are all Poisson with the following probability functions, respectively, 
\begin{align*}
    &  P(N_{i,l}=n\mid C_l=j)= p(n;\Tilde{\lambda}_l^{(j)}(\boldsymbol{x}_i)),\\
    &  P(N_{i,l}^{\text{r}}=n\mid C_l=j)= p(n;\mu_l^{(j)}(\boldsymbol{x}_i)),\\
    & P(N_{i,l}^{\text{ibnr}}=n\mid C_l=j)= p(n;\nu_l^{(j)}(\boldsymbol{x}_i)),
\end{align*}
where 
\begin{align*}
    & \Tilde{\lambda}_l^{(j)}(\boldsymbol{x}_i)=e_{i,l}\lambda^{(j)}(\boldsymbol{x}_i)(d_{l}-d_{l-1})\\    
    & \mu_l^{(j)}(\boldsymbol{x}_i) = e_{i,l}\left(\int_{d_{l-1}}^{d_{l}}F_{U\mid t,\boldsymbol{x}_i}(\tau-t)\,dt\right)\lambda^{(j)}(\boldsymbol{x}_i),\\
    & \nu_l^{(j)}(\boldsymbol{x}_i) =e_{i,l} \left(\int_{d_{l-1}}^{d_{l}}(1-F_{U\mid t,\boldsymbol{x}_i}(\tau-t))\,dt\right)\lambda^{(j)}(\boldsymbol{x}_i),
\end{align*}
and 
$$p(n;\lambda)=\frac{\lambda^n e^{-\lambda}}{n!}.$$
As a result, for $l=1,2,\dots,$ $N_{i,l}$, $N_{i,l}^\text{r}$, and $N_{i,l}^{\text{ibnr}}$ follow mixed Poisson distributions with the probability functions, respectively,
\begin{align*}
    & P(N_{i,l}=n) = \sum_{j=1}^g \pi_{lj}p(n;\Tilde{\lambda}_l^{(j)}(\boldsymbol{x}_i)),\\
    & P(N_{i,l}^{\text{r}}=n) = \sum_{j=1}^g \pi_{lj}p(n;\mu_l^{(j)}(\boldsymbol{x}_i)),\\
    & P(N_{i,l}^{\text{ibnr}}=n) = \sum_{j=1}^g \pi_{lj}p(n;\nu_l^{(j)}(\boldsymbol{x}_i)).
\end{align*}
\hfill $\square$  
\end{proposition} 
The above proposition is straightforward given our assumption of constant intensity functions given a realization of the hidden state. When $C_l = j$, the intensity functions for the claim arrival process, reported claim process, and IBNR claim process become constant over the period $[d_{l-1},d_l)$, resulting in Poisson processes for the claim arrival process and time-varying Poisson processes for the reported and IBNR claim processes. It is also evident that the three processes share the same hidden parameter process. This observation leads directly to the conclusion that all three observed processes are Poisson-HMMs.

While Poisson-HMMs have traditionally been employed to model claim count processes for entire portfolios, our application extends this methodology to the individual level (or other granular unit). Properties similar to those in \cite{badescu2016marked} can be easily derived, but we refrain from showing them in the manuscript, out of considerations related to the manuscript's length.

\section{Likelihood of the Observed Data} \label{sec:likelihood}
In this section, we propose the discrete-time version of our model, both with and without the Dirichlet assumption for the reporting probability vector. The motivation for switching to a discrete-time framework arises from the complexities associated with the continuous-time framework, particularly the difficulties in maximizing the likelihood of the observed data. We begin by outlining the likelihood of the observed data in the continuous-time framework and briefly discussing these complexities. This will serve as a motivation for our discrete-time alternative.

\subsection{Continuous-time Model} \label{subsec:cont_likelihood}

To estimate the IBNR reserve, we seek to maximize the likelihood of the observed data. Let $n_{i,l}^\text{r}$ denotes the number of claims incurred by policyholder (or any granular unit) $i$ during period $[d_{l-1},d_l]$ and are reported by $d_T = \tau$. These claims, ordered by their arrival time, are described through the three random variables $\{t_{ijl},u_{ijl},\boldsymbol{z}_{ijl}:i=1,\dots,m;j=1,\dots,n_{i,l}^\text{r}\}$, where $t_{ijl}$ is the claim arrival time of the $j$-th ordered claim for policyholder $i$ in period $[d_{l-1},d_l)$, $u_{ijl}$ is its reporting delay, and $\boldsymbol{z}_{ijl}$ is its development information. Our set of observed data is thus given by: $$\mathcal{O}_C = \{t_{ijl},u_{ijl},\boldsymbol{z}_{ijl}:i=1,\dots,m;j=1,\dots,n_{i,l}^\text{r};l=1,\dots,T\}.$$ Let $\boldsymbol{N}_{i}^{(\text{r},s:t)}  = (N_{i,s}^\text{r},\dots,N_{i,t}^\text{r})$ be the discretely observed reported claim process from period $s$ to period $t$ for policyholder $i$, and $\boldsymbol{n}_{i}^{(\text{r},s:t)}=(n_{i,s}^\text{r},\dots,n_{i,t}^\text{r})$ be its realization, it is easy to show that the likelihood of the observed data is composed of three terms (e.g., see \cite{antonio2014micro} and \cite{fung2021new}):

\begin{equation} \label{likelihood}
\begin{split}
\mathcal{L}^{(1)}(\boldsymbol{\Phi}_1\mid \mathcal{O}_C) \propto &\underbrace{P\left(\boldsymbol{N}_1^{(\text{r},1:T)}=\boldsymbol{n}_1^{(\text{r},1:T)},\dots,\boldsymbol{N}_m^{(\text{r},1:T)}=\boldsymbol{n}_m^{(\text{r},1:T)}\right)}_{\text{First Term (Claim Frequency)}} \times \underbrace{\prod_{i=1}^m \prod_{l=1}^T \prod_{j=1}^{n_{i,l}^\text{r}} \frac{f_{U\mid t_{ijl},\boldsymbol{x}_i}(u_{ijl})}{F_{U\mid t_{ijl},\boldsymbol{x}_i}(\tau-t_{ijl})}}_{\text{Second Term (Reporting Delay)}} \\
& \times \underbrace{\prod_{i=1}^m \prod_{l=1}^T \prod_{j=1}^{n_{i,l}^\text{r}} f_{\boldsymbol{Z}\mid u_{ijl},t_{ijl},\boldsymbol{x}_i}(\boldsymbol{z}_{ijl})}_{\text{Third Term (Claim Development)}},
\end{split}
\end{equation}
where \(\boldsymbol{\Phi}_1\) represents the vector of the model's parameters, including: the initial distribution \(\boldsymbol{\pi}_1\) and the transition matrix \(\boldsymbol{\Gamma}\) of the HMM, the regression coefficients \(\boldsymbol{\theta}_j\) for the \(\lambda^{(j)}\)'s, and the parameters associated with modeling the reporting delay and claim severity. 

The ``Claim Frequency" component captures the probability of observing the reported claim counts across all policyholders and time periods. It characterizes the underlying claim frequency patterns in the data.

The ``Reporting Delay" component accounts for the right-truncated reporting delay for claims that are reported before the valuation date $\tau$. It is modeled as a regression dependent on the policy characteristics and the time of occurrence. Fitting the right-truncated reporting delay regression requires special attention and can be a challenging task because: random right-truncation is not as well studied in survival analysis as left-truncation, and empirical data for reporting delay show the necessity of having a flexible mixture regression model, which is not straight-forward for right-truncated data.

The ``Claim Development" component is concerned with the development process after a claim is reported and involves right-censored data. In case we need to estimate the IBNR reserve, we will need to model claim severity, and thus, this third term is represented as $p_{Y\mid U,t,\boldsymbol{x}}(y)$ where $Y$ denotes the claim severity. This models the relationship between claim severity and the reporting delay $U$, considering the policy characteristics $\boldsymbol{x}$.  

As discussed in the introduction, maximizing the above likelihood is a challenging task. We observe that the third term, representing the claim development $\boldsymbol{z}$, is independent of the parameters in the first two terms. Hence, it can be maximized separately. From Proposition \ref{prop2}, we know that the reported claim counts' distribution depends on the reporting delay, making the maximization of the first two terms of $\mathcal{L}^{(1)}$ a formidable task. As previously mentioned, the common approach in the micro-level reserving literature is to employ a two-step maximization. This two-step maximization approach rests on the assumption of independence between claim frequency and reporting delay, which is not a realistic assumption. 

In what follows, we drop the claim development component from our analysis as we focus on the interaction between frequency and delay.   

\subsection{Discrete-time Models} \label{subsec:disc_likelihood}
\subsubsection{The Multinomial Model}
With the aim of joint modeling of frequency and delay, we deviate from the continuous-time framework to a discrete-time framework by aggregating the events towards a coarser time scale. To this end, we assume that claims can be reported with a maximum delay of $D$ periods. This means that if a claim happens at period $t$ (i.e. the claim happens in $[d_t,d_{t+1})$), it can be reported in periods $t+d$ ($d=0,1,\dots,D$). We also, for simplicity, assume that the predetermined time points $d_t$ $(t=0,\dots,T)$ are equidistant, so that $d_{t+1}-d_{t}=1$ and $d_T = \tau$. We denote the probability that a claim from policyholder with risk attributes $\boldsymbol{x}_i$ that occurred in period $t$ is reported in period $t+d$ by $p_{t}(d;\boldsymbol{x}_i)$. It is easy to show that the discretely observed claim arrival process, $\{N_{i,1},N_{i,2},\dots,N_{i,T} \}$, the discretely observed reported claim process, $\{N_{i,1}^{\text{r}},\dots,N_{i,T}^{\text{r}} \}$, and the discretely observed IBNR claim process, $\{N_{i,1}^{\text{ibnr}},\dots,N_{i,T}^{\text{ibnr}}\}$, fall under the class of Poisson-HMMs, with the same hidden parameter process $\{C_1,C_2,\dots\}$ as $N(t;\boldsymbol{x}_i)$, and with state-dependent Poisson intensities given by:
\begin{align*}
    \Tilde{\lambda}_t^{(j)}(\boldsymbol{x}_i)&=e_{i,t}\lambda^{(j)}(\boldsymbol{x}_i), \\
    \mu_t^{(j)}(\boldsymbol{x}_i) & = e_{i,t}\lambda^{(j)}(\boldsymbol{x}_i)p_{i,t}^\text{r}, \\
    \nu_t^{(j)}(\boldsymbol{x}_i) & =e_{i,t} \lambda^{(j)}(\boldsymbol{x}_i)(1-p_{i,t}^\text{r}),
\end{align*}
respectively, where $p_{i,t}^\text{r} = \sum_{d=0}^{\min\{D,T-t\}}p_{t}(d;\boldsymbol{x}_i)$ is the probability that a claim from policyholder $i$ that occurred in period $t$ is reported before the reserve valuation date.

Let $Z_{i,t,d}$ denote the number of claims from policyholder $i$ that occurred in period $t$ and are reported in period $t+d$. Note that $N_{i,t}^\text{r} = \sum_{d=0}^{\min\{D,T-t\}}Z_{i,t,d}$. Our observed data is now given by: $$\mathcal{O}_D = \{n_{i,t}^{\text{r}},z_{i,t,d}\mid i=1,\dots,m;t = 1,\dots,T;d=0,\dots,D; t+d\le T \},$$ where $n_{i,t}^\text{r}$ and $z_{i,t,d}$ are the realizations of $N_{i,t}^\text{r}$ and $Z_{i,t,d}$, respectively.  

Similar to \cite{verbelen2022modeling}, we make the assumption that given $N_{i,t}^\text{r}$, the counts $Z_{i,t,d}$ $(d=0,\dots,\min\{D,T-t\})$ follow a multinomial distribution with probabilities $\frac{p_{t}(d;\boldsymbol{x}_i)}{p_{i,t}^\text{r}}$ $(d=0,\dots,\min\{D,T-t\})$. Note that for $t \le T-D$, all claims occurred in period $t$ is fully observed, i.e., $N_{i,t}^\text{r} = N_{i,t}$, and the counts $(Z_{i,t,0},\dots,Z_{i,t,D})\mid N_{i,t}^\text{r}\sim \text{Multinomial}(N_{i,t}^\text{r},\boldsymbol{p}_t(\boldsymbol{x}_i)),$ where $\boldsymbol{p}_t(\boldsymbol{x}_i)=(p_t(0;\boldsymbol{x}_i),\dots,p_t(D;\boldsymbol{x}_i))$ .

The likelihood of the observed data is then given by:
\begin{equation} 
    \begin{split}
    \mathcal{L}^{(2)}(\boldsymbol{\Phi}_2\mid \mathcal{O}_D) & \propto  P\left(\boldsymbol{N}_1^{(\text{r},1:T)}=\boldsymbol{n}_1^{(\text{r},1:T)},\dots,\boldsymbol{N}_m^{(\text{r},1:T)}=\boldsymbol{n}_m^{(\text{r},1:T)}\right) \\ &\times \prod_{i=1}^m \prod_{t=1}^T \frac{n_{i,t}^\text{r}!}{\prod_{d=0}^{\min\{D,T-t\}}z_{i,t,d}!} \prod_{d=0}^{\min\{D,T-t\}} \left(\frac{p_{t}(d;\boldsymbol{x}_i)}{p_{i,t}^\text{r}}\right)^{z_{i,t,d}},
    \end{split}
\end{equation}
where \(\boldsymbol{\Phi}_2\) represents the vector of the model's parameters, including: the initial distribution \(\boldsymbol{\pi}_1\) and the transition matrix \(\boldsymbol{\Gamma}\) of the HMM, the regression coefficients \(\boldsymbol{\theta}_j\) for the \(\lambda^{(j)}\)'s, and the parameters related to the modeling of \(\boldsymbol{p}_t(\boldsymbol{x}_i)\), denoted by \(\boldsymbol{\delta}\). An equivalent way to write the multinomial component of the likelihood above is to decompose it into a series of conditional binomial likelihoods. That is, we can write $\mathcal{L}^{(2)}$ as:
\begin{equation} 
    \begin{split}
    \mathcal{L}^{(2)}(\boldsymbol{\Phi}_2\mid \mathcal{O}_D) & \propto  P\left(\boldsymbol{N}_1^{(\text{r},1:T)}=\boldsymbol{n}_1^{(\text{r},1:T)},\dots,\boldsymbol{N}_m^{(\text{r},1:T)}=\boldsymbol{n}_m^{(\text{r},1:T)}\right) \\ &\times \prod_{i=1}^m \prod_{t=1}^T \prod_{d=1}^{\min\{D,T-t\}} \binom{\sum_{j=0}^{d}z_{i,t,j}}{z_{i,t,d}} q_t(d;\boldsymbol{x}_i)^{z_{i,t,d}}(1-q_t(d;\boldsymbol{x}_i))^{\sum_{j=0}^{d-1}z_{i,t,d}},
    \end{split}
\end{equation}
where $q_t(d;\boldsymbol{x}_i) = \frac{p_t(d;\boldsymbol{x}_i)}{\sum_{j=0}^{d}p_t(j;\boldsymbol{x}_i)}$ represents the conditional probability that a claim incurred at period $t$ by policyholder with risk attributes $\boldsymbol{x}_i$ is reported in period $t+d$, given that the claim is reported by period $t+d$. 
Thus, we break down the multinomial likelihood by considering the binomial likelihood of observing each count given the cumulative counts and adjusting the success probabilities accordingly.  Note that $p_t(d;\boldsymbol{x}_i)$ can be obtained iteratively from $q_t(d;\boldsymbol{x}_i)$ by:
\begin{align*}
    & p_t(D;\boldsymbol{x}_i) = q_t(D;\boldsymbol{x}_i) \\
    & p_t(d;\boldsymbol{x}_i) = q_t(d;\boldsymbol{x}_i)\times \left(1-\sum_{j=d+1}^D p_t(j;\boldsymbol{x}_i)\right) \quad (d=D-1,\dots,1) \\
    & p_t(0;\boldsymbol{x}_i) = 1 - \sum_{j=1}^D p_t(j;\boldsymbol{x}_i).
\end{align*}

\subsubsection{The Dirichlet-Multinomial Model}
The probabilities of reporting delay often exhibit temporal variability, which can be characterized as either explainable or unexplainable. Explainable variations in reporting delay are systematic variations that arise from differences in the period of occurrence of claims and/or the risk attributes associated with policyholders. For example, when working with daily data, claims occurring on weekends may exhibit different reporting patterns compared to those occurring on weekdays. Similarly, the month in which a claim occurs can also influence reporting delay probabilities. In contrast, unexplainable variability refers to the unsystematic fluctuations in reporting delay that are not captured by our covariates. In cases where data exhibits high random variability in reporting delay, both continuous-time and multinomial models may produce interval estimates for IBNR claim counts that are significantly off from the actual IBNR claim count, as we will observe in our data analysis. While an exact cause for the high fluctuations cannot be precisely identified without additional data collection, we believe that there are unobserved covariates who have direct and interaction effects on the delay probabilities. 

To this end, we add randomness to the reporting delay probabilities $p_{t}(d;\boldsymbol{x}_i)$'s by assuming that $\boldsymbol{p}_t(\boldsymbol{x}_i) = (p_{t}(0;\boldsymbol{x}_i),\dots,p_{t}(D;\boldsymbol{x}_i))$ follows a Dirichlet distribution with parameters $\boldsymbol{\eta}_{i,t} = (\eta_{t,0}(\boldsymbol{x}_i),\dots,\eta_{t,D}(\boldsymbol{x}_i))$. This choice is motivated by the fact that the Dirichlet distribution serves as the conjugate prior for the multinomial distribution, which significantly simplifies the mathematical treatment and estimation of our model. 

The parameters \(\boldsymbol{\eta}_{i,t} = (\eta_{t,0}(\boldsymbol{x}_i), \dots, \eta_{t,D}(\boldsymbol{x}_i))\) can be specified in various ways. One approach is to model \(\boldsymbol{\eta}_{i,t}\) using a regression with fixed effects, where these parameters are expressed as a function of observed covariates, such as policyholder characteristics or temporal factors. Alternatively, we can incorporate the Dirichlet assumption as a random effect. For instance, we could assume that for each period \(t\), the reporting delay probability vector \(\boldsymbol{p}_t\) follows a Dirichlet distribution with parameters \(\boldsymbol{\eta}_t\). This formulation represents a "global model," where the reporting delay probability vector at period \(t\) for each policyholder \(i\) is drawn from this global distribution. However, this model assumes that all policyholders have similar reporting patterns, which may not always hold in practice. To refine this approach further, we could cluster the policyholders into groups \(\mathcal{I}_1, \dots, \mathcal{I}_{n'}\), and assume that for each period \(t\) and for each group \(\mathcal{I}\), the reporting delay probability vector \(\boldsymbol{p}_t^{\mathcal{I}}\) follows a Dirichlet distribution with parameters \(\boldsymbol{\eta}_t^{\mathcal{I}}\). In this case, the reporting delay probability vector at period \(t\) for all policyholders \(i \in \mathcal{I}\) would be drawn from this group-specific model, allowing for a more granular modeling of heterogeneity in reporting behaviors across different groups. Lastly, a random coefficients model could be employed, where both the intercept and the coefficients associated with the covariates in the regression model are allowed to vary randomly across different clusters or observations (in this case, different policies). This approach provides further flexibility, enabling us to account for substantial random fluctuations in the delay probabilities.

We make use of the following proposition which can also be found in \cite{lawless1994adjustments}:

\begin{proposition}
    If $\boldsymbol{p}_t(\boldsymbol{x}_i) = (p_{t}(0;\boldsymbol{x}_i),\dots,p_{t}(D;\boldsymbol{x}_i))$ follows a Dirichlet distribution with parameters $\boldsymbol{\eta}_{i,t} = (\eta_{t,0}(\boldsymbol{x}_i),\dots,\eta_{t,D}(\boldsymbol{x}_i))$, then the conditional probabilities $q_t(d;\boldsymbol{x}_i)= \frac{p_t(d;\boldsymbol{x}_i)}{\sum_{j=0}^{d}p_t(j;\boldsymbol{x}_i)}$ $(d=1,\dots,D)$ are independently distributed beta random variables with 
    $$q_t(d;\boldsymbol{x}_i) \sim \text{Beta}(\eta_{t,d}(\boldsymbol{x}_i),\sum_{j=0}^{d-1}\eta_{t,j}(\boldsymbol{x}_i)).$$
\end{proposition}

Thus, the binomial likelihoods components in $\mathcal{L}^{(2)}$ would be replaced with beta-binomial likelihoods components, and the likelihood for our model with the Dirichlet assumption becomes:

\begin{equation} 
    \begin{split}
    \mathcal{L}^{(3)}(\boldsymbol{\Phi}_3\mid \mathcal{O}_D) & \propto  P\left(\boldsymbol{N}_1^{(\text{r},1:T)}=\boldsymbol{n}_1^{(\text{r},1:T)},\dots,\boldsymbol{N}_m^{(\text{r},1:T)}=\boldsymbol{n}_m^{(\text{r},1:T)}\right) \\ &\times \prod_{i=1}^m \prod_{t=1}^T \prod_{d=1}^{\min\{D,T-t\}} \binom{\sum_{j=0}^{d}z_{i,t,j}}{z_{i,t,d}}\frac{B(z_{i,t,d}+\eta_{t,d}(\boldsymbol{x}_i),\sum_{j=0}^{d-1}(z_{i,t,d}+\eta_{t,j}(\boldsymbol{x}_i)))}{B(\eta_{t,d}(\boldsymbol{x}_i),\sum_{j=0}^{d-1}\eta_{t,j}(\boldsymbol{x}_i))},
    \end{split}
\end{equation}
where $B$ is the beta function, and $\boldsymbol{\Phi}_3$ represents the vector of the model's parameters, including: the initial distribution \(\boldsymbol{\pi}_1\) and the transition matrix \(\boldsymbol{\Gamma}\) of the HMM, the regression coefficients \(\boldsymbol{\theta}_j\) for the \(\lambda^{(j)}\)'s, and the parameters related to the modeling of the \(\boldsymbol{\eta}\)'s, denoted by \(\boldsymbol{\delta}\).   

\textbf{Remark:} With \( g=1 \) (single state), we have that \( N_{i,t}^\text{r} \) follows a Poisson distribution, and so, this Dirichlet assumption can be seen as an extension of the model of \cite{verbelen2022modeling}. We provide the EM algorithm for this model in the next section.

\section{Parameter Estimation} \label{sec:est}

In this section, we present the estimation methodology employed to fit the discrete-time models from the previous section.  To achieve this, we utilize the EM algorithm, providing an iterative framework for updating the model parameters to maximize the likelihood of the observed reported claims while considering the unobserved hidden states of the HMM. The EM algorithm for fitting the continuous-time model using the two-step maximization can be found in Appendix \ref{appendixA}.

The EM algorithm for the Multinomial model extends the algorithm outlined in \cite{verbelen2022modeling}, where the reported claims process now follows a Poisson-HMM, rather than a Poisson process. On the other hand, the EM algorithm for the Dirichlet-Multinomial model requires special attention. This is because the Dirichlet assumption introduces additional complexity in estimating the probability vector, making the E-step analytically intractable. Therefore, we will need to switch from the EM algorithm to the Monte Carlo EM (MCEM) algorithm, which uses Monte Carlo simulations to approximate the E-step and handle the complexities introduced by the Dirichlet distribution.

\subsection{The Multinomial Model} \label{em_muli}

Recall that the likelihood of the observed data for the discrete-time Multinomial model, $\mathcal{L}^{(2)}$, is given by:
\begin{equation*} 
    \begin{split}
    \mathcal{L}^{(2)}(\boldsymbol{\Phi}_2\mid \mathcal{O}_D) & \propto  P\left(\boldsymbol{N}_1^{(\text{r},1:T)}=\boldsymbol{n}_1^{(\text{r},1:T)},\dots,\boldsymbol{N}_m^{(\text{r},1:T)}=\boldsymbol{n}_m^{(\text{r},1:T)}\right) \\ &\times \prod_{i=1}^m \prod_{t=1}^T \frac{n_{i,t}^\text{r}!}{\prod_{d=0}^{\min\{D,T-t\}}z_{i,t,d}!} \prod_{d=0}^{\min\{D,T-t\}} \left(\frac{p_{t}(d;\boldsymbol{x}_i)}{p_{i,t}^\text{r}}\right)^{z_{i,t,d}},
    \end{split}
\end{equation*}
where $\boldsymbol{N}_{i}^{(r,1:T)}  = (N_{i,1}^r,\dots,N_{i,T}^r)$ denote the discretely observed reported claim process from period 1 to $T$ for policy $i$, and $\boldsymbol{n}_{i}^{(r,1:T)}=(n_{i,1}^r,\dots,n_{i,T}^r)$ is its realization. As we established in Proposition \ref{prop2}, our model dictates that the discretely observed reported claims process for policy $i$, $\{N_{i,1}^\text{r},\dots,N_{i,k}^\text{r}\}$, originates from the class of Poisson-HMM. Specifically, we demonstrated that:
$$N_{i,t}^\text{r}\mid C_t=j \sim \text{Poisson}(e_{i,t}\lambda^{(j)}(\boldsymbol{x}_i)p_{i,t}^\text{r}).$$

The likelihood for the ``Claim Frequency'' component is thus given by:
$$P\left(\boldsymbol{N}_{1}^{(r,1:T)}=\boldsymbol{n}_{1}^{(r,1:T)},\dots,\boldsymbol{N}_{m}^{(r,1:T)}=\boldsymbol{n}_{m}^{(r,1:T)}\right)=\boldsymbol{\pi}_1\boldsymbol{P}_1(\boldsymbol{n}_1^r)\boldsymbol{\Gamma}\boldsymbol{P}_2(\boldsymbol{n}_2^r)\dots\boldsymbol{\Gamma}\boldsymbol{P}_T(\boldsymbol{n}_T^r)\mathbf{1}^T=:L_T,$$
where 
$$\boldsymbol{n}_{t}^\text{r}  =(n_{1,t}^\text{r},\dots,n_{m,t}^\text{r}), $$  $$\boldsymbol{P}_t(\boldsymbol{n}_t^r) = \text{diag}\left\{\prod_{i=1}^m \boldsymbol{P}(N_{i,t}^r=n_{i,t}^r\mid C_t=1),\dots,\prod_{i=1}^m \boldsymbol{P}(N_{i,t}^r=n_{i,t}^r\mid C_t=g)\right\},$$
and $\mathbf{1}$ is a row vector of 1s. 

\subsubsection*{The Forward-Backward Algorithm} 
Before sketching the EM algorithm, we first define the forward and backward probabilities as follows:
\begin{align*}
    & \alpha_{tj} := P(\boldsymbol{N}_{1}^{(r,1:t)}=\boldsymbol{n}_{1}^{(r,1:t)},\dots,\boldsymbol{N}_{m}^{(r,1:t)}=\boldsymbol{n}_{m}^{(r,1:t)}, C_t=j), \\
    & \beta_{tj} := P(\boldsymbol{N}_{1}^{(r,t+1:T)}=\boldsymbol{n}_{1}^{(r,t+1:T)},\dots,\boldsymbol{N}_{m}^{(r,t+1:T)}=\boldsymbol{n}_{m}^{(r,t+1:T)}\mid  C_t=j),
\end{align*}
respectively. These probabilities are used to iteratively update the hidden state probabilities during the EM algorithm, and they have the following properties (see \cite{macdonald2016hidden} for details):
\begin{align*}
    & P(C_t=j\mid \boldsymbol{N}_{1}^{(r,1:T)}=\boldsymbol{n}_{1}^{(r,1:T)},\dots,\boldsymbol{N}_{m}^{(r,1:T)}=\boldsymbol{n}_{m}^{(r,1:T)}) = \frac{\alpha_{tj}\beta_{tj}}{L_T}, \\
    & P(C_{t-1}=j,C_t=k\mid \boldsymbol{N}_{1}^{(r,1:T)}=\boldsymbol{n}_{1}^{(r,1:T)},\dots,\boldsymbol{N}_{m}^{(r,1:T)}=\boldsymbol{n}_{m}^{(r,1:T)}) = \frac{\alpha_{t-1,j}\gamma_{jk}\mathbb{P}_k(\boldsymbol{n}_t^r)\beta_{tk}}{L_T},
\end{align*}
where $\mathbb{P}_k(\boldsymbol{n}_t^r)=\prod_{i=1}^m \boldsymbol{P}(N_{i,t}^r=n_{i,t}^r\mid C_t=k)$. 

They are computed using the following recursive relations:
\begin{align*}
    &  \boldsymbol{\alpha}_1 = \boldsymbol{\pi}_1\boldsymbol{P}_1(\boldsymbol{n}_1^r),\\
    & \boldsymbol{\alpha}_t=\boldsymbol{\alpha}_{t-1}\boldsymbol{\Gamma}\boldsymbol{P}_t(\boldsymbol{n}_t^r),\quad t=2,\dots,T,\\
    & \boldsymbol{\beta}_T = \mathbf{1}, \\ 
    & \boldsymbol{\beta}_t=\boldsymbol{\Gamma}\boldsymbol{P}_{t+1}(\boldsymbol{n}_{t+1}^r) \boldsymbol{\beta}_{t+1},\quad t=T-1,T-2,\dots,1.
\end{align*}

We now proceed with the details of the EM algorithm. 

\subsubsection*{Complete Data Likelihood} 
The EM algorithm iteratively estimates the parameters $\boldsymbol{\Phi}_2$, where the number of hidden states, $g$, is treated as preset. The algorithm consists of two main steps: the E-step and the M-step, which will be demonstrated in the following subsections. We first introduce the following notations: 
\begin{itemize}
    \item $\boldsymbol{c}^{(T)}=(c_1,\dots,c_T)$: the unobserved states of the HMM,
    \item $u_{tj} := \mathbbm{1}\{C_t = j\}$, and
    \item $v_{tjk} := \mathbbm{1}\{C_{t-1}=j, C_{t}=k\}$.
\end{itemize}
 
We let our complete data consist of all claims that occurred before our reserve valuation date (both reported and unreported), as well as the realizations of the states of the HMM. Thus, the complete data is given by:
\begin{align*}
    \mathcal{C} &= \mathcal{O}_D \cup \{n_{i,t}^{\text{ibnr}}, z_{i,t,d} \mid i=1, \dots, m; \, t=1, \dots, T; \,d=1,\dots,D\, t+d>T\} \cup \{c_1, \dots, c_T\} \\ 
    &= \{n_{i,t}, z_{i,t,d}, c_t \mid i=1, \dots, m; \, t=1, \dots, T; \, d=0, \dots, D\}.
\end{align*}
Note that the complete data assumes that we know the total number of claims that occurred in each period for each policyholder, as well as the periods in which these claims are reported.
The likelihood of the complete data is then given by:
\begin{equation*} 
\begin{split}
    \mathcal{L}^{(2)}_{\text{C}}(\boldsymbol{\Phi}_2\mid \mathcal{C}) 
 & =\pi_{1,c_1}\prod_{t=2}^T\gamma_{c_{t-1},c_t}\prod_{t=1}^T \prod_{i=1}^m \mathbb{P}_{c_t}(n_{i,t}) \\  
 & \times \prod_{i=1}^m \prod_{t=1}^T \prod_{d=1}^{D} \binom{\sum_{j=0}^{d}z_{i,t,j}}{z_{i,t,d}} q_t(d;\boldsymbol{x}_i)^{z_{i,t,d}}(1-q_t(d;\boldsymbol{x}_i))^{\sum_{j=0}^{d-1}z_{i,t,d}},
\end{split}
\end{equation*}
where $\mathbb{P}_{c_t}(n_{i,t}) = P(N_{i,t}=n_{i,t}\mid C_t = c_t)$.

It is straightforward to see that we can write the log-likelihood of the complete data as:
\begin{align*}
     l_{\text{C}}^{(2)}(\boldsymbol{\Phi}_2\mid \mathcal{C}) & = \sum_{j=1}^g u_{1j}\log \pi_{1j} +\sum_{j=1}^g\sum_{k=1}^g\left(\sum_{t=2}^Tv_{tjk}\right)\log \gamma_{jk} + \sum_{j=1}^g\sum_{t=1}^T \sum_{i=1}^m u_{tj}\log \mathbb{P}_{j}(n_{i,t}) \\
    & + \sum_{i=1}^m \sum_{t=1}^T \sum_{d=1}^D \left[\log \binom{\sum_{j=0}^{d}z_{i,t,j}}{z_{i,t,d}} + z_{i,t,d}\log q_t(d;\boldsymbol{x}_i) + \left(\sum_{j=0}^{d-1}z_{i,t,d}\right)\log(1-q_t(d;\boldsymbol{x}_i))\right]. 
\end{align*}
We now outline the E-step and M-step of the EM algorithm that are applied to maximize the complete-data log-likelihood. 

\subsubsection*{E-step} 
In the $k$-th E-step, we compute the conditional expectation of the complete-data log-likelihood given the observed data and the current estimator $\boldsymbol{\Phi}_2^{(k-1)}$ of the model parameters $\boldsymbol{\Phi}_2$. Note that the unobserved components of our complete data consist of the hidden states, $c^{(T)}$, as well as the IBNR claims, $\{n_{i,t}^{\text{ibnr}}, z_{i,t,d} \mid i=1,\dots,m; t=1,\dots,T; t+d > T\}$. The expectation of these components given the observed data and $\boldsymbol{\Phi}_2^{(k-1)}$ should thus be derived. The conditional expectation is given by:

\begin{equation} \label{equ:estepL2}
    \begin{split}
    \mathbb{E}[l_{\text{C}}^{(2)}(\boldsymbol{\Phi}_2 \mid \mathcal{C}) \mid \boldsymbol{\Phi}_2^{(k-1)}, \mathcal{O}_D ] & = \sum_{j=1}^g \hat{u}_{1j}^{(k)} \log \pi_{1j} + \sum_{j=1}^g \sum_{k=1}^g \left(\sum_{t=2}^T \hat{v}_{tjk}^{(k)}\right) \log \gamma_{jk} \\ 
    & + \sum_{j=1}^g \sum_{t=1}^T \sum_{i=1}^m \hat{u}_{tj}^{(k)} \left[\hat{n}_{i,t,j}^{(k)} \log(\lambda^{(j)}(\boldsymbol{x}_i)) - \lambda^{(j)}(\boldsymbol{x}_i)\right] \\
    & + \sum_{i=1}^m \sum_{t=1}^T \sum_{d=1}^D \left[\hat{z}_{i,t,d}^{(k)} \log q_t(d; \boldsymbol{x}_i) + \left(\sum_{j=0}^{d-1} \hat{z}_{i,t,d}^{(k)}\right) \log(1-q_t(d; \boldsymbol{x}_i))\right] \\ 
    & + \text{constant}, 
    \end{split}
\end{equation}

where the expressions for $\hat{u}_{tj}^{(k)}$ and $\hat{v}_{tjl}^{(k)}$ are given by:

\begin{equation} \label{eqn:uhat}
    \begin{split}
    \hat{u}_{tj}^{(k)} & = \mathbb{E}\left(u_{tj} \mid \boldsymbol{n}^{(r,1:T)}, \boldsymbol{\Phi}_2^{(k-1)}\right) \\
    & = \mathbb{P}\left(C_t = j \mid \boldsymbol{n}^{(r,1:T)}, \boldsymbol{\Phi}_2^{(k-1)}\right) \\
    & = \frac{\alpha_{tj}^{(k-1)} \beta_{tj}^{(k-1)}}{L_T^{(k-1)}}, 
    \end{split}
\end{equation}

and

\begin{equation} \label{eqn:vhat}
    \begin{split}
    \hat{v}_{tjl}^{(k)} & = \mathbb{E}\left(v_{tjl} \mid \boldsymbol{n}^{(r,1:T)}, \boldsymbol{\Phi}_2^{(k-1)}\right) \\
    & = \mathbb{P}\left(C_{t-1} = j, C_t = l \mid \boldsymbol{n}^{(r,1:T)}, \boldsymbol{\Phi}_2^{(k-1)}\right) \\
    & = \frac{\alpha_{t-1,j}^{(k-1)} \gamma_{jl}^{(k-1)} \prod_{i=1}^{m} P(N_{i,t}^\text{r} = n_{i,t}^\text{r} \mid C_t = l, \boldsymbol{\Phi}_2^{(k-1)}) \beta_{tl}^{(k-1)}}{L_T^{(k-1)}}
    \end{split}
\end{equation}

In these expressions, the forward probabilities, backward probabilities, and observed likelihood are obtained using the model parameters $\boldsymbol{\Phi}_2 = \boldsymbol{\Phi}_2^{(k-1)}$ from the $(k-1)$-th iteration. The terms $\hat{u}_{tj}^{(k)}$ and $\hat{v}_{tjl}^{(k)}$ represent the updated probabilities of being in state $j$ at time $t$ and the transition probability from state $j$ to state $l$ at time $t$, respectively, based on the current model parameter estimates $\boldsymbol{\Phi}_2^{(k-1)}$ and the observed data.

The expressions for $\hat{n}_{i,t,j}^{(k)}$ and $\hat{z}_{i,t,d}^{(k)}$ are given by:
\begin{align*}
    & \hat{n}_{i,t,j}^{(k)} := \mathbb{E}(N_{i,t} \mid C_t = j, \mathcal{O}_D, \boldsymbol{\Phi}_2^{(k-1)}) = n_{i,t}^\text{r} + \mathbbm{1}\{t > T-D\} e_{i,t} \lambda_j^{(k-1)}(\boldsymbol{x}_i) \sum_{d=T-t+1}^D p_{t}^{(k-1)}(d; \boldsymbol{x}_i), \\
    & \hat{z}_{i,t,d}^{(k)} := \mathbb{E}(Z_{i,t,d} \mid \mathcal{O}_D, \boldsymbol{\Phi}_2^{(k-1)}) = \begin{cases} 
      z_{i,t,d}, & t \le T-D \\
      \left(\sum_{j=1}^g \pi_{tj}^{(k-1)} e_{i,t} \lambda_j^{(k-1)}(\boldsymbol{x}_i) \right) p_{t}^{(k-1)}(d; \boldsymbol{x}_i), & t > T-D
   \end{cases}
\end{align*}
where $p_t^{(k-1)}(d; \boldsymbol{x}_i)$ and $\lambda_j^{(k-1)}(\boldsymbol{x}_i)$ are the estimates of $p_t(d; \boldsymbol{x}_i)$ and $\lambda^{(j)}(\boldsymbol{x}_i)$, respectively, computed using the parameter estimates obtained from the M-step in the $(k-1)$-th iteration. The term $\hat{n}_{i,t,j}^{(k)}$ represents the updated expectation of the total number of claims that occurred in period $t$ for policyholder $i$, given that the state of the HMM is $j$, while the term $\hat{z}_{i,t,d}^{(k)}$ represents the updated expectation of the portion of these claims that are reported in period $t+d$. Note that all claims that occurred at $t \le T-D$ are reported and are part of the observed data $\mathcal{O}_D$, and thus, the expected values of $N_{i,t}$ and $Z_{i,t,d}$ ($t \le T-D; d=0, \dots, D$) given the observed data are equal to $n_{i,t}$ and $z_{i,t,d}$, respectively.

By maximizing this expected complete-data log-likelihood, we can iteratively update the model parameters in the M-step of the algorithm, leading to improved parameter estimates that better fit the observed data.

\subsubsection*{M-step} 
The M-step in the EM algorithm involves maximizing the expected complete-data log-likelihood (Equation (\ref{equ:estepL2})) with respect to the model parameters $\boldsymbol{\Phi}_2$, subject to the constraints 
\begin{align*}
    &  \sum_{j=1}^g\pi_{1j}=1, \text{ and} \\
    & \sum_{l=1}^g\gamma_{jl}=1,\quad j=1,\dots,g.
\end{align*}
For this step, we can obtain the updated estimates for $\boldsymbol{\pi}_1$ and $\boldsymbol{\Gamma}$ separately by maximizing the first two terms of Equation (\ref{equ:estepL2}), respectively. The optimal values for $\pi_{1j}^{(k)}$ and $\gamma_{jl}^{(k)}$ at the $k$th iteration are given by:

\begin{align} 
&\pi_{1j}^{(k)} = \hat{u}_{1j}^{(k)},\quad j=1,\dots,g, \label{eqn:pi}\\
& \gamma_{jl}^{(k)} = \frac{\sum_{t=2}^T \hat{v}_{tjl}^{(k)}}{\sum_{l=1}^g\sum_{t=2}^T \hat{v}_{tjl}^{(k)}},\quad j,l=1,\dots,g. \label{eqn:gamma}
\end{align}

As for the third term:
$$\sum_{t=1}^T \sum_{i=1}^m \hat{u}_{tj}^{(k)}\left[\hat{n}_{i,t,j}^{(k)}\log(\lambda^{(j)}(\boldsymbol{x}_i)) - \lambda^{(j)}(\boldsymbol{x}_i)\right],$$
it is easy to see that for each $j$, this is a weighted Poisson log-likelihood, with the exception that $\hat{n}_{i,t,j}^{(k)}$ can be non-integer. Yet, we can still obtain the updated regression parameters for each $\lambda^{(j)}$ using quasi-likelihood methods available through standard packages in R. Finally, for the term:$$\sum_{i=1}^m \sum_{t=1}^T \left[\hat{z}_{i,t,d}^{(k)}\log q_t(d;\boldsymbol{x}_i) + \left(\sum_{j=0}^{d-1}\hat{z}_{i,t,d}^{(k)}\right)\log(1-q_t(d;\boldsymbol{x}_i))\right],$$
it is easy to see that for each $d$, this is a Binomial log-likelihood, again with the exception that $\hat{z}_{i,t,d}^{(k)}$ can be non-integer. For each $d$, we model $q_t(d;\boldsymbol{x}_i)$ as a regression function of the policy-holder risk attributes, $\boldsymbol{x}_i$, and time-dependent covariates (e.g. the month of the period $t$). The Binomial log-likelihood can then be optimized using quasi-likelihood methods available through standard package in R. 

The model parameters are then given by $\boldsymbol{\Phi}_2 = \{\boldsymbol{\pi}_1,\boldsymbol{\Gamma},\boldsymbol{\theta}_1,\dots,\boldsymbol{\theta}_g,\boldsymbol{\delta}_1,\dots,\boldsymbol{\delta}_D\}$, where $\boldsymbol{\theta}_j$ is the vector of regression parameters for $\lambda^{(j)}(\boldsymbol{x})$, and $\boldsymbol{\delta}_d$ is the vector of regression parameters for $q_t(d;\boldsymbol{x}).$

\subsubsection*{Convergence} 
We continue iterating the E-step and M-step until the relative distance between the estimates from consecutive iterations falls below a predetermined threshold. The relative distance is defined as
\begin{align*}
    d(\boldsymbol{\Phi}_2^{(k-1)},\boldsymbol{\Phi}_2^{(k)}) & = \sum_{j=1}^g \left| \frac{\pi_{1j}^{(k-1)}-\pi_{1j}^{(k)}}{\pi_{1j}^{(k-1)}} \right| + \sum_{j=1}^g \sum_{l=1}^g\left|\frac{\gamma_{jl}^{(k-1)}-\gamma_{jl}^{(k)}}{\gamma_{jl}^{(k-1)}} \right| + \sum_{j=1}^g \sum_{s=1}^{||\boldsymbol{\theta}_j||_0}\left|\frac{\theta_{js}^{(k-1)}-\theta_{js}^{(k)}}{\theta_{js}^{(k-1)}}\right|\\&+\sum_{d=1}^D \sum_{s=1}^{||\boldsymbol{\delta}_j||_0}\left|\frac{\delta_{js}^{(k-1)}-\delta_{js}^{(k)}}{\delta_{js}^{(k-1)}}\right|.
\end{align*}
The EM algorithm converges when the relative distance is below the threshold.

\subsubsection*{Initialization of Model Parameters} 
To effectively apply the EM algorithm, appropriate initialization of model parameters is crucial. Note that we have complete data (all claims are reported) for periods $t \le T - D$, so we can fit the model for such data to obtain parameter estimates as initialization 

\subsubsection*{Model Selection} 
Finally, we need to decide on the number of states \( g \). Typically, a larger \( g \) would provide a better fit, but it could also lead to overfitting. We usually select the model with the lowest chosen information criterion (e.g., AIC or BIC). For fast and efficient computation, we can employ a backward fitting strategy similar to \cite{badescu2019marked}. We start by fitting the model with a large \( g \), and then iteratively delete a state until the chosen information criterion stops decreasing. In this strategy, we use the parameter estimates obtained from fitting with \( g \) states to initialize the algorithm for \( g-1 \) states, with the initial values for \( \boldsymbol{\pi}_1 \) and \( \boldsymbol{\Gamma} \) being normalized estimates excluding the deleted state. This approach should make the estimation process very efficient.

\subsubsection*{Under-/Overflow Problems} 
As noted by \cite{macdonald2016hidden}, the computation of forward and backward probabilities, along with the subsequent calculations in the E- and M- steps, is susceptible to under- or overflow issues. \cite{macdonald2016hidden} recommends scaling the forward and backward probability computations by taking the logarithms of these quantities and provides code solutions for this purpose. However, for our model, additional considerations are needed.

The computation of forward and backward probabilities involves matrix multiplication of the matrices $\boldsymbol{P}_t(\boldsymbol{n}_t^r)$. Unlike normal HMMs fitted to single time series, where these matrices are diagonal matrices containing the probability of the count process obtaining its observed value given the different states, our model's probabilities are the product of the observed count process for each individual. For our model, this matrix is defined as
$$ \boldsymbol{P}_t(\boldsymbol{n}_t^r) = \text{diag}\left\{\prod_{i=1}^m \boldsymbol{P}(N_{i,t}^r=n_{i,t}^r\mid C_t=1),\dots,\prod_{i=1}^m \boldsymbol{P}(N_{i,t}^r=n_{i,t}^r\mid C_t=g)\right\}.$$

It becomes evident that the elements of the diagonal matrix are extremely small, potentially causing underflow problems in computing forward and backward probabilities, as well as subsequent calculations, even after scaling. To address this issue, we implement the \texttt{LogSumExp} algorithm (e.g., see \cite{blanchard2021accurately}), which is effective in mitigating numerical instability arising from the small probabilities. 

\subsection{The Dirichlet-Multinomial Model} \label{em_dri}
\subsubsection*{Complete Data Likelihood}
We now aim to find the parameter estimates that maximize $\mathcal{L}^{(3)}$ using the EM algorithm. As we illustrated, the Dirichlet-Multinomial model differs from the Multinomial model by the assumption that the probability vector $\boldsymbol{p}_{t}(\boldsymbol{x}_i)$ is random. To this end, we let our complete data be given by:
\begin{align*}
    \tilde{\mathcal{C}} &= \mathcal{C} \cup \{p_{i,t,d} \mid i=1,\dots,m; t=1,\dots,T; d=1,\dots,D\} \\ 
    &= \{n_{i,t}, z_{i,t,d}, p_{i,t,d}, c_t \mid t=1,\dots,T; d=0,\dots,D\},
\end{align*}
where $p_{i,t,d}$ is the realization of $p_d(t; \boldsymbol{x}_i)$. That is, we now assume we know the actual reporting probability vector $\boldsymbol{p}_t(\boldsymbol{x}_i)$. The likelihood of the complete data is then given by:
\begin{equation*} 
\begin{split}
    \mathcal{L}^{(3)}_{\text{C}}(\boldsymbol{\Phi}_3\mid \tilde{\mathcal{C}}) 
 & =\pi_{1,c_1}\prod_{t=2}^T\gamma_{c_{t-1},c_t}\prod_{t=1}^T \prod_{i=1}^m \mathbb{P}_{c_t}(n_{i,t}) \\  
 & \times \prod_{i=1}^m \prod_{t=1}^T \left(\underbrace{\frac{n_{i,t}!}{\prod_{d=0}^D z_{i,t,d}!}\prod_{d=1}^{D} p_{i,t,d}^{z_{i,t,d}}}_{\text{Multinomial Distribution}} \times \underbrace{\frac{1}{B(\boldsymbol{\eta}_{i,t})}\prod_{d=0}^D p_{i,t,d}^{\eta_{t,d}(\boldsymbol{x}_i)-1}}_{\text{Dirichlet Distribution}}\right),
\end{split}
\end{equation*}
where $\mathbb{P}_{c_t}(n_{i,t}) = P(N_{i,t}=n_{i,t}\mid C_t = c_t)$. Note that the ``Multinomial Component" in the complete data likelihood is constant since it does not depend on the model's parameters. The log-likelihood of the complete data is then given by:
\begin{align*}
     l_{\text{C}}^{(3)}(\boldsymbol{\Phi}_3\mid \tilde{\mathcal{C}}) & = \sum_{j=1}^g u_{1j}\log \pi_{1j} +\sum_{j=1}^g\sum_{k=1}^g\left(\sum_{t=2}^Tv_{tjk}\right)\log \gamma_{jk} + \sum_{j=1}^g\sum_{t=1}^T \sum_{i=1}^m u_{tj}\log \mathbb{P}_{j}(n_{i,t}) \\
    & + \sum_{i=1}^m \sum_{t=1}^T -\log B(\boldsymbol{\eta}_{i,t}) \sum_{d=1}^D (\eta_{t,d}(\boldsymbol{x}_i)-1)\log p_{i,t,d}+ \text{constant}. 
\end{align*}

\subsubsection*{E-Step} 
In the $k$-th iteration of the E-step, we take the expectation of $l_{\text{C}}^{(3)}(\boldsymbol{\Phi}_3\mid \tilde{\mathcal{C}})$ with respect to $\boldsymbol{\Phi}_3^{(k-1)}$ and $\mathcal{O}_D$. We have:
\begin{equation} \label{eq:estep_dir}
    \begin{split}
    \mathbb{E}\left[l_{\text{C}}^{(3)}(\boldsymbol{\Phi}_3 \mid \tilde{\mathcal{C}}) \mid \boldsymbol{\Phi}_3^{(k-1)}, \mathcal{O}_D \right] &= \sum_{j=1}^g \hat{u}_{1j}^{(k)} \log \pi_{1j} + \sum_{j=1}^g \sum_{l=1}^g \left(\sum_{t=2}^T \hat{v}_{tjl}^{(k)}\right) \log \gamma_{jl} \\
    & \quad + \sum_{j=1}^g \sum_{t=1}^T \sum_{i=1}^m \hat{u}_{tj}^{(k)} \left[\hat{n}_{i,t,j}^{(k)} \log (\lambda^{(j)}(\boldsymbol{x}_i)) - \lambda^{(j)}(\boldsymbol{x}_i) \right] \\
    & \quad + \sum_{i=1}^m \sum_{t=1}^T - \log B(\boldsymbol{\eta}_{i,t}) \sum_{d=1}^D (\eta_{t,d}(\boldsymbol{x}_i) - 1) \widehat{\log p_{i,t,d}}^{(k)} \\
    & \quad + \text{constant}, 
    \end{split}
\end{equation}

where $\hat{u}_{tj}^{(k)}$ and $\hat{v}_{tjl}^{(k)}$ are given by Equations (\ref{eqn:uhat}) and (\ref{eqn:vhat}), respectively, and 
\begin{align*}
\hat{n}_{i,t,j}^{(k)} &:= \mathbb{E}\left[N_{i,t} \mid C_t = j, \mathcal{O}_D, \boldsymbol{\Phi}_3^{(k-1)}\right] \\
&= n_{i,t}^\text{r} + \mathbbm{1}\{t > T - D\} e_{i,t} \lambda_j^{(k-1)}(\boldsymbol{x}_i) \mathbb{E}\left[\sum_{d=T-t+1}^D p_{t}(d; \boldsymbol{x}_i) \mid C_t = j, \mathcal{O}_D, \boldsymbol{\Phi}_3^{(k-1)}\right], \\
\widehat{\log p_{i,t,d}}^{(k)} &:= \mathbb{E}\left[\log p_{t}(d; \boldsymbol{x}_i) \mid \mathcal{O}_D, \boldsymbol{\Phi}_3^{(k-1)}\right],
\end{align*}

where $\lambda_j^{(k-1)}(\boldsymbol{x}_i)$ is the estimate of $\lambda^{(j)}(\boldsymbol{x}_i)$ computed using the parameter estimates obtained from the M-step in the $(k-1)$-th iteration. We do not have closed-form analytical expressions for $\hat{n}_{i,t,j}^{(k)}$ ($t > T-D$) and $\widehat{\log p_{i,t,d}}^{(k)}$, and thus, numerical methods are needed to compute these expectations. We can obtain Monte Carlo estimates of these expectations using samples from the conditional distribution $\boldsymbol{p}_t(\boldsymbol{x}_i) = (p_t(0; \boldsymbol{x}_i), \dots, p_t(D; \boldsymbol{x}_i))$, using the parameters $\boldsymbol{\Phi}_3$. This, however, adds an extra layer of challenge because there is no guarantee that the likelihood will increase with each iteration, unlike in the case of the normal EM algorithm, as we are replacing the actual expectation with a numerical approximation.

\textbf{Remark:} The computation of the forward probabilities, backward probabilities, the likelihood, and consequently, $\hat{u}_{tj}$ and $\hat{v}_{tjk}$ in the E-step differs slightly for the two models considered. The differences arise when computing the probability \( P(N_{i,t}^\text{r} = n_{i,t}^\text{r} \mid  \mathcal{O}_D, \boldsymbol{\Phi}^{(k-1)}) \). As previously discussed, this probability depends on the reporting delay. For the discrete-time Multinomial model, the estimation of the probability vector \( \boldsymbol{p}_t(\boldsymbol{x}_i) \) changes after each iteration of the algorithm, and so we use \( \boldsymbol{p}_t(\boldsymbol{x}_i) \) estimated from the \( k-1 \)-th iteration in our computation in the \( k \)-th iteration. As for the Dirichlet-Multinomial model, \( \boldsymbol{p}_t(\boldsymbol{x}_i) \) is a Dirichlet random vector, and thus we compute 
\begin{align*}
    & P(N_{i,t}^\text{r} = n_{i,t}^\text{r} \mid  \mathcal{O}_D, \boldsymbol{\Phi}_3^{(k-1)}) = \\
    & \int_{\boldsymbol{p}} P(N_{i,t}^\text{r} = n_{i,t}^\text{r} \mid  \boldsymbol{p}_t(\boldsymbol{x}_i) = \boldsymbol{p}, \mathcal{O}_D, \boldsymbol{\Phi}_3^{(k-1)}) \times P(\boldsymbol{p}_t(\boldsymbol{x}_i) = \boldsymbol{p} \mid  \mathcal{O}_D, \boldsymbol{\Phi}_3^{(k-1)}) \, d\boldsymbol{p}.
\end{align*}

\subsubsection*{Distribution of $\boldsymbol{p}_t \mid \mathcal{O}_D$} 
In the case where \( t \le T - D \), we have:
\[
\boldsymbol{p}_t(\boldsymbol{x}_i) \mid \mathcal{O}_D \sim \text{Dirichlet}(\eta_{t,0}(\boldsymbol{x}_i) + z_{i,t,0}, \dots, \eta_{t,D}(\boldsymbol{x}_i) + z_{i,t,D}),
\]
since the Dirichlet distribution is a conjugate prior of the multinomial distribution. Challenges arise in the case where \( t > T - D \), as we do not observe \( z_{i,t,d} \) for \( d > T - t \), and thus we lose the nice conjugate property. Therefore, we need to derive the conditional distribution of \(\boldsymbol{p}_t\) for \( t > T - D \). 

The probability of $\boldsymbol{p}_t(\boldsymbol{x}_i)$ equal $\boldsymbol{p}=(p_0,\dots,p_D)$ given the observed data is given by:
\begin{align*}
    P(\boldsymbol{p}_t(\boldsymbol{x}_i) = \boldsymbol{p}\mid \mathcal{O}_D) & \propto P(\mathcal{O}_D\mid \boldsymbol{p}_t(\boldsymbol{x}_i)=\boldsymbol{p}) \times P(\boldsymbol{p}_t(\boldsymbol{x}_i) = \boldsymbol{p}), 
\end{align*}
where
\begin{align*}
    & P(\mathcal{O}_D\mid \boldsymbol{p}_t(\boldsymbol{x}_i)=\boldsymbol{p})  \propto P(N_{i,t}^\text{r}=n_{i,t}^\text{r}\mid \boldsymbol{p}_t(\boldsymbol{x}_i)=\boldsymbol{p})\times\\&P(Z_{i,t,0}=z_{i,t,0},\dots,Z_{i,t,T-t}=z_{i,t,T-t}\mid N_{i,t}^\text{r}=n_{i,t}^\text{r},\boldsymbol{p}_t(\boldsymbol{x}_i)=\boldsymbol{p}).
\end{align*}
Thus, 
\begin{align*}
    & P(\boldsymbol{p}_t(\boldsymbol{x}_i)  = \boldsymbol{p}\mid \mathcal{O}_D) 
     \propto  P(\boldsymbol{p}_t(\boldsymbol{x}_i) = \boldsymbol{p}) \times P(N_{i,t}^\text{r}=n_{i,t}^\text{r}\mid \boldsymbol{p}_t(\boldsymbol{x}_i)=\boldsymbol{p}) \times \\ &  P(Z_{i,t,0}=z_{i,t,0},\dots,Z_{i,t,T-t}=z_{i,t,T-t}\mid N_{i,t}^\text{r}=n_{i,t}^\text{r},\boldsymbol{p}_t(\boldsymbol{x}_i)=\boldsymbol{p})  \\
    & \propto \prod_{d=0}^D (p_d)^{\eta_{t,d}(\boldsymbol{x}_i)-1} \times \left(\sum_{j=1}^g \pi_{tj}\frac{e^{-\lambda^{(j)}(\boldsymbol{x}_i)p^\text{r}}(\lambda^{(j)}(\boldsymbol{x}_i)p^\text{r})^{n_{i,t}^\text{r}}}{n_{i,t}^\text{r}!}\right)\times\frac{n_{i,t}^\text{r}!}{\prod_{d=0}^{T-t}z_{i,t,d}!} \prod_{d=0}^{T-t}\left(\frac{p_d}{p^\text{r}}\right)^{z_{i,t,d}} \\
    & \propto \underbrace{\left(\prod_{d=0}^{T-t} (p_d)^{z_{i,t,d}+\eta_{t,d}(\boldsymbol{x}_i)-1}\right) \times \left( \prod_{d=T-t+1}^D (p_d)^{\eta_{t,d}(\boldsymbol{x}_i)-1}\right)}_{h} \times \underbrace{\sum_{j=1}^g \pi_{tj}\left(\lambda^{(j)}(\boldsymbol{x}_i)\right)^{n_{i,t}^\text{r}}e^{-\lambda^{(j)}(\boldsymbol{x}_i)p^\text{r}}}_{g},
\end{align*}
where $p^{\text{r}}=\sum_{d=0}^{T-t}p_d$. Note that the conditional distribution in this case is not a standard distribution, and thus, we cannot sample from it directly.

\subsubsection*{Rejection Sampling} 
Samples from the conditional distribution of $\boldsymbol{p}_t(\boldsymbol{x}_i)$ can be selected by multivariate rejection sampling. We can write $P(\boldsymbol{p}_t(\boldsymbol{x}_i)=p\mid \mathcal{O}_D)=ah(\boldsymbol{p})g(\boldsymbol{p}).$ It is straightforward to see that $h$ is the density function of Dirichlet distribution with parameters $(z_{i,t,0}+\eta_{t,0}(\boldsymbol{x}_i),\dots,z_{i,t,T-t}+\eta_{t,T-t}(\boldsymbol{x}_i),\eta_{t,T-t+1}(\boldsymbol{x}_i),\dots,\eta_{t,D}(\boldsymbol{x}_i)),$ and thus, we can easily sample from it. We perform the rejection sampling as follows:
\begin{itemize}
    \item \textit{Step 1:} sample $\boldsymbol{p}$ from $h$, and independently, sample $u$ from a Uniform(0,1) distribution.
    \item \textit{Step 2:} If $u \le \frac{g(\boldsymbol{p})}{\sup_{\boldsymbol{p}}\{g(\boldsymbol{p})\}},$ then accept $\boldsymbol{p}$. If not, go to \textit{Step 1}.
\end{itemize}

Note that $\sup_{\boldsymbol{p}}\{g(\boldsymbol{p})\}\le \sum_{j=1}^g \pi_{tj}(\lambda^{(j)}(\boldsymbol{x}_i))^{n_{i,t}^\text{r}}$. By obtaining a sample for $\boldsymbol{p}_t(\boldsymbol{x}_i)$ given $\mathcal{O}_D$ and $\boldsymbol{\Phi}_3^{(k-1)},$ we can estimate $\hat{n}_{i,t,j}^{(k)}$ and $\widehat{\log p_{i,t,d}}^{(k)}$ using Monte Carlo methods. 

\subsubsection*{M-Step} 
We now maximize Equation (\ref{eq:estep_dir}) from the E-step after replacing $\hat{n}_{i,t,j}^{(k)}$ ($t > T-D$) and $\widehat{\log p_{i,t,d}}^{(k)}$ with their Monte Carlo estimates. Again, the updated estimates of $\boldsymbol{\pi}_1$ and $\boldsymbol{\Gamma}$ are given by Equations (\ref{eqn:pi}) and (\ref{eqn:gamma}), respectively. Moreover, similar to the Multinomial model case, given each $j$, it is straightforward to see that the third term is a weighted Poisson log-likelihood, and thus can be fit using standard packages in R. Finally, for the fourth and final term, it is straightforward to see that this is a Dirichlet log-likelihood. As previously illustrated, there are several flexible ways to model \(\boldsymbol{\eta}\). For our data analysis, we model \(\boldsymbol{\eta}\) as a regression function of \(\boldsymbol{x}_i\) and time-dependent covariates. We obtain the parameter estimates for this Dirichlet regression using the \texttt{DirichletReg} package in R, which rescales the probability estimates from the E-step to ensure that the constraints are satisfied.

\subsubsection*{Convergence Criterion, Model Initialization \& Model Selection} 

We deviate from the standard EM algorithm and use the MCEM algorithm, as we compute the expectations in the E-step using Monte Carlo methods. In the MCEM algorithm, it is not guaranteed that the likelihood will increase after each iteration, unlike the standard EM algorithm (see \cite{booth1999maximizing}). Thus, special care should be taken. Many methods have been proposed for approximating the EM algorithm (see \cite{ruth2024review} for a review). From a practical point of view, we use the same convergence criterion as we have for the previous model (the relative distance criterion), as this would have a negligible effect on the parameter estimates even though the likelihood can fluctuate. 

For model initialization, we fit the model to our complete data up to period \( T - D \) (as we did in the Multinomial case), and use the parameter estimates as the initialization for the algorithm. Finally, we choose the number of states \( g \) using a backward strategy as explained before.

\section{IBNR Reserve Prediction} \label{sec:ibnr}

The estimation of the IBNR claim reserve varies slightly across the three models. Below, we outline the steps for the continuous-time model:

\textbf{IBNR Claim Reserve Estimation Steps} 
\begin{enumerate}[leftmargin=*]
    \item \textit{Fit Right-Truncated Reporting Delay Regression}: Estimate the parameters of the right-truncated reporting delay regression model to obtain the fitted distribution of the reporting delay \(F_{U\mid t,\boldsymbol{x}}(u)\).

    \item \textit{Compute Reporting Delay Probability}: Use the fitted reporting delay regression to compute the probability \(\int_{d_{l-1}}^{d_l} F_{U\mid t,\boldsymbol{x}_i}(\tau-t) \, dt\) for each policyholder \(i=1,\dots,m\) and each period \(l=1,\dots,T\). This probability is used in maximizing the likelihood of the discretely observed reported claims processes and simulating the IBNR claim counts.

    \item \textit{Fit the Discretely Observed Reported Claim Process}: Maximize the likelihood of the observed reported claim counts using the EM algorithm. This involves estimating the initial state probabilities and the transition matrix for the HMM, as well as the claim frequency regression parameters.

    \item \textit{Viterbi's Global Decoding}: Apply Viterbi's algorithm to find the most likely sequence of hidden states \(\boldsymbol{c}^*\) given the observed reported claim counts \(\boldsymbol{n}^{(r,1:T)}\) and the fitted model parameters from step 3.

    \item \textit{Simulate IBNR Claim Counts}: For each policyholder \(i=1,\dots,m\) and each period \(l=1,\dots,T\) where the exposure is positive, simulate the IBNR claim count \(N_{i,l}^{\text{IBNR}}\) using the Poisson distribution with mean \(\lambda_l^{(c_l^*)}(\boldsymbol{x}_i) \times (1 - \int_{d_{l-1}}^{d_l} F_{U\mid t,\boldsymbol{x}_i}(\tau-t) \, dt)\) for the corresponding hidden state from the previous step. The total IBNR claim count at time \(\tau\) is then given by \(\sum_{i=1}^m \sum_{t=1}^T \hat{N}_{i,l}^{\text{IBNR}}\), where \(\hat{N}_{i,l}^{\text{IBNR}}\) is the simulated IBNR claim count for policyholder \(i\) and period \(l\).

    \item \textit{Estimate the Reserve}: For the continuous-time model, simulate the occurrence time \(t\) for each IBNR claim from the previous step, assuming it is uniformly distributed across the period of occurrence. Then, simulate the reporting delay \(u\) from the fitted regression model for reporting delay, given that \(u > \tau - t\). Finally, simulate the claim severity using a fitted regression model for claim severity, given the policyholder characteristics \(\boldsymbol{x}_i\), the occurrence time \(t\), and the simulated reporting delay \(u\). The sum of these simulated amounts will constitute the reserve.
\end{enumerate}

Note that \textit{Steps 5} \& \textit{6} are repeated 1000 times to obtain a distribution for the IBNR claim counts and IBNR reserves. The point estimate for these amounts is the mean of the 1000 simulated values.

The process differs slightly for the discrete-time models. For the Multinomial model, the reporting delay component is fitted simultaneously with the frequency component using the EM algorithm. Thus, the estimation procedure starts from \textit{Step 3}. The steps of estimating the IBNR claim counts (\textit{Steps 4} \& \textit{5} from above) are essentially the same as in the continuous-time model, but we use the estimated reporting probability vector $\boldsymbol{p}_t(\boldsymbol{x}_i)$ in the simulation instead of the continuous distribution $F_U$. For the Dirichlet model, we obtain estimates for $\boldsymbol{\eta}_{i,t}$, the Dirichlet parameters of $\boldsymbol{p}_t(\boldsymbol{x}_i)$, in \textit{Step 3}. This is used to first simulate $\boldsymbol{p}_t(\boldsymbol{x}_i)$, which is then used in \textit{Step 5} to estimate the IBNR claim count as in the case of the Multinomial model. We can simulate the period of reporting $d$ using the estimated reporting probability vector $\boldsymbol{p}_t(\boldsymbol{x}_i)$ and then use it to estimate the IBNR reserve by simulating from a fitted model for the average claim severity given the policyholder characteristics $\boldsymbol{x}_i$, the occurrence time $t$, and the simulated period of reporting $d$.

\section{Case Study: Real-life Data} \label{sec:data}

In this section, we apply the proposed models to estimate the IBNR claim count for a major European auto insurance company. We then compare these estimates with those obtained from the traditional Chain Ladder method. This analysis aims to evaluate the performance of our models in a practical insurance context. Additionally, we investigate how the joint modeling of occurrence and reporting in the discrete-time models performs compared to the continuous-time model, where we maximize the likelihood of the observed data using a two-step maximization process. This is particularly interesting given the loss of information when moving to discrete-time models.

\subsection{Data Analysis}

\subsubsection*{Data Description}
Our dataset contains information on Physical Damage policies issued by the company between January 2009 and December 2017. During these years, a total of 293,709 one-year policies were issued, resulting in 106,187 reported claims. For each policy, the dataset includes the policy start and end dates, as well as various vehicle and policyholder attributes. These attributes include the age of the driver, the age of the car, the class of the car (A, B, or C), the fuel type (Gasoline or Diesel), whether the contract is a renewal or issued to a new policyholder, and the region (five regions). For each claim, the dataset records the day on which the claim occurred, the day on which the claim was reported, and the progression of claim payments. In this study, we focus on estimating the IBNR claim count, which requires us to concentrate on the time of occurrence and reporting delay of each claim. Therefore, we will not consider the payments in our analysis.

\subsubsection*{Reporting Delay}
From the 106,197 claims reported between January 2009 and December 2017, only two were reported more than two years after the occurrence date. This suggests that nearly all claims occurring before 2016 have been reported. Table \ref{table:reporting_delay} displays the distribution of reporting delays in months for these claims. A delay of zero months indicates a claim was reported in the same month it occurred, while a delay of one month means it was reported the following month, and so on. The majority of claims (81.71\%) were reported in the same month they occurred. Over 99\% of claims were reported within five months, and only 0.1\% were reported after ten months. This indicates that most claims are reported promptly after occurrence, with only a small fraction experiencing delays beyond ten months.

\begin{table}[h] 
\centering
\small
\setlength{\extrarowheight}{2pt} 
\begin{tabularx}{\textwidth}{|>{\centering\arraybackslash}p{2.2cm}|*{11}{>{\centering\arraybackslash}X|}}
\hline
\textbf{Delay (months)} & 0      & 1      & 2      & 3      & 4      & 5      & 6      & 7      & 8      & 9      & 10+     \\ \hline
\textbf{No. of claims}  & 66475 & 11617  & 1580   & 642    & 342    & 192    & 141    & 105    & 80    & 52     & 130     \\ \hline
\textbf{\% of total}    & 81.71 & 14.28 & 1.94  & 0.79  & 0.42  & 0.24  & 0.17  & 0.13  & 0.10  & 0.06  & 0.16   \\ \hline
\textbf{Cumulative \%}  & 81.71 & 95.99 & 97.93 & 98.72 & 99.14 & 99.38 & 99.55 & 99.68 & 99.78 & 99.84 & 100.0 \\ \hline
\end{tabularx}
\caption{Reporting Delay for Claims Occurring Before 2016}
\label{table:reporting_delay}
\end{table}

\subsubsection*{Frequency}
\begin{figure}[h]
    \includegraphics[width=0.5\textwidth]{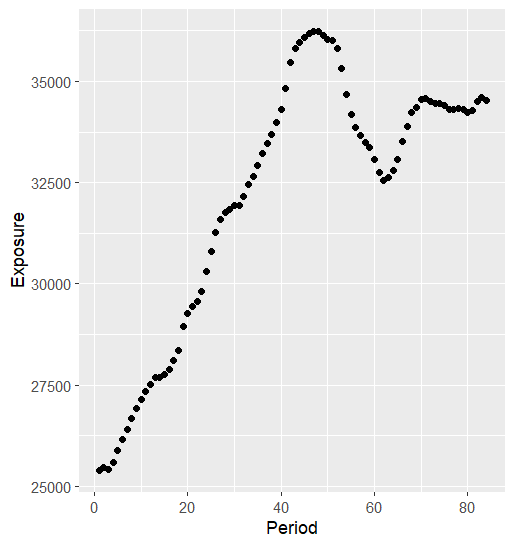}
    \includegraphics[width=0.5\textwidth]{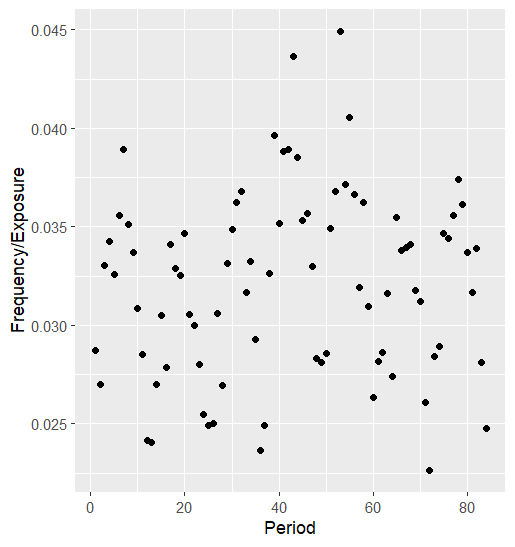}
    \caption{Exposure \& Claim Count Per Exposure Between Jan 2009 and Dec 2015}
    \label{fig:freqexp}
\end{figure}

Figure \ref{fig:freqexp} illustrates the trends in exposure (left) and claim count per exposure (right) from January 2009 to December 2015. The exposure, which represents the number of policies at risk, exhibited a steady increase from January 2009 until December 2012. Following this period, it decreased consistently until January 2015, after which it began to rise again. In contrast, the claim count per exposure displayed significant fluctuations over the same period. It reached a minimum of 0.023 claims per exposure in December 2014 and peaked at 0.045 claims per exposure in May 2013. This variation in claim frequency highlights the need for a sophisticated model capable of capturing such dynamic changes. The HMM employed in our model is well-suited for this purpose, as it can account for the underlying state changes that drive fluctuations in claim frequency over time.

\subsection{IBNR Claim Count Prediction} 
In this subsection, we outline the four different models used to estimate the IBNR claim count for our data, along with their specifications. The models considered are the Continuous-time model, the Multinomial model, the Dirichlet-Multinomial model, and the traditional Chain Ladder method.

For the discrete-time models, we set the maximum reporting delay, $D$, to 9 months. This choice is based on the observation that nearly 99.9\% of claims are reported within this period, ensuring that the majority of claims are captured within the model. Additionally, the choice of $D=9$ is motivated by the limitations of the \texttt{DirichletReg} function in R, which encounters numerical instability with larger values due to the exceedingly small probabilities associated with longer delays.

\textbf{Specifications for the Fitted Models}
\begin{itemize}[leftmargin=*]
    \item \textit{The Continuous-time Model (CM):} As previously explained, our approach begins by fitting the distribution for the reporting delay, which is then utilized to fit our model, and consequently estimate the IBNR claim count. A notable challenge in this process is the right truncation of the reporting delay, which complicates the fitting of the distribution. We provide the continuous-time model with an advantage over the discrete-time models as we fit the reporting delay using all claims that occurred before the reserve valuation date $\tau$, including those that have not yet been reported. This overcomes the issue of right truncation, allowing for a more accurate fit.
    
    While this method is not feasible in practice — since unreported claims cannot be known beforehand — it serves to highlight the strengths of the discrete-time models, which we will demonstrate in the results section. In a practical scenario, one could use a ``back censoring" technique, where a time threshold within the training period is identified. Claims occurring before this threshold would have been reported by the valuation date, allowing the model to be fit without concern for right truncation.
    
    In our analysis, the log-logistic proportional hazard regression performed well in fitting the reporting delay data, as evidenced by the Cox-Snell residuals plot in Figure \ref{CoxSnell}. The covariates used in the regression include the policyholder-level covariates mentioned in the data description subsection and time-dependent covariates such as the month in which the claim occurred and the day of occurrence.

    \begin{figure}[h]
        \centering
        \includegraphics[width=0.6\textwidth]{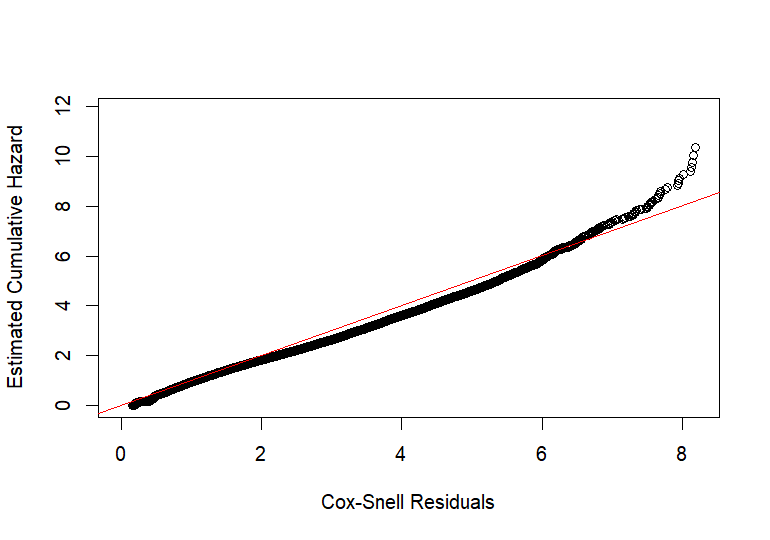}
        \caption{Cox-Snell Residual Plot for Assessing the Fit of the Reporting Delay Model}
        \label{CoxSnell}
    \end{figure}

    \item \textit{The Multinomial Model (MM):} In Section \ref{em_muli}, we outlined the EM algorithm required to fit the Multinomial Model, where we model the Binomial probabilities \( q_t(d;\boldsymbol{x}_i) = \frac{p_t(d;\boldsymbol{x}_i)}{\sum_{j=0}^d p_t(j;\boldsymbol{x}_i)} \) for each \( d \in \{1,\dots,D\} \) as a regression function of the policyholder's risk attributes \(\boldsymbol{x}_i\) and time-dependent covariates. The policyholder-level covariates are those described in the data section, while the time-dependent covariates include the month of occurrence and the weekday on which the period ends, both of which were found to be significant in our regression analysis. Given that \( q_t(d;\boldsymbol{x}_i) \) is small for \( d > 1 \), we use a complementary log-log link function to model it. For \( d = 1 \), we use a standard log link function, as is typical for Binomial GLMs. We fit the Binomial component of the likelihood using the standard \texttt{glm} function in R.

    \item \textit{The Dirichlet-Multinomial Model (DM):} The fitting process for the Dirichlet-Multinomial model is thoroughly detailed in Section \ref{em_dri}. We model the Dirichlet prarameters \(\boldsymbol{\eta}\) as a regression function of the same covariates used for \(q_t(d;\boldsymbol{x}_i)\) in the Multinomial Model. 

    \item \textit{The Chain Ladder (CL):} We estimate the IBNR claim count using the Chain Ladder method on data aggregated monthly, serving as the baseline for comparing our models. As with the discrete-time models, we assume that claims are reported no later than 9 months after the month of occurrence.
\end{itemize}

\textbf{Remarks}
\begin{itemize}[leftmargin=*]
    \item While setting \( D = 9 \) months simplifies the model, it may lead to underestimating the IBNR claim count because it ignores claims that are reported after this period. However, since nearly all claims are reported within this window, the number of missed claims should be very small. In practice, practitioners can estimate this remaining number empirically using the formula:

\[
\sum_{t=1}^{T} \sum_{d=T-t+1}^{\infty} \hat{z}_{t,d}
\]

where $T$ is such that $d_T = \tau$, \( \hat{z}_{t,d} \) is the estimated number of claims that occurred in period \( t \) and were reported after a delay of \( d \) periods, and is calculated as:

\[
\hat{z}_{t,d} = \frac{1}{T-1} \sum_{l=1}^{t-1} z_{l,d}
\]

Here, \( z_{l,d} \) represents the number of claims that occurred in period \( l \) and were reported with a delay of \( d \) periods. Essentially, \( \hat{z}_{t,d} \) is the average number of claims reported after a delay of \( d \) periods, based on the observed claims data from previous periods.

\item For the continuous-time model, we do not assume a maximum reporting delay, so by following the steps in Section \ref{sec:ibnr}, we can estimate the full number of IBNR claims, unlike the discrete-time models. However, for a fair comparison with the discrete-time models, we use the estimated parameters after fitting the continuous-time model to estimate only the IBNR claims with a maximum delay period of 9 months after the month of claim occurrence. This is achieved by adjusting \textit{Step 5} so that the probability factor \(\left(1 - \int_{d_{l-1}}^{d_l} F_{U\mid t,\boldsymbol{x}_i}(\tau-t) \, dt\right)\) is replaced with:

\[
\int_{d_{l-1}}^{d_l} P(U+t \in [\tau, t_{\text{max}}]\mid t,\boldsymbol{x}_i) \, dt,
\]

where \(t\) is the day of claim occurrence, \(U\) is the reporting delay random variable, and \(t_{\text{max}}\) is the date by which the claim should be reported if we assume a maximum delay of 9 months after the month of claim occurrence. For example, if the claim occurrence date is in January, then \(t_{\text{max}}\) will be the end of October for the same year. This adjustment ensures that only the first 9 months of reporting delay are considered in the estimation, aligning the continuous-time model's output with the assumptions made in the discrete-time models.
\end{itemize}

\subsection{Results} 
We present the results of estimating the IBNR claim count using the four different models, focusing on claims reported within 9 months after the month of occurrence. Estimations were performed at the end of each month from January 2014 to December 2016, resulting in 36 valuation points. Since our dataset extends to December 2017, and given that more than 99.99\% of claims are reported within a year of occurrence, we have a reliable actual IBNR claim count to compare against our model estimates.

\begin{figure}[t]
    \centering
    \includegraphics[width=\textwidth]{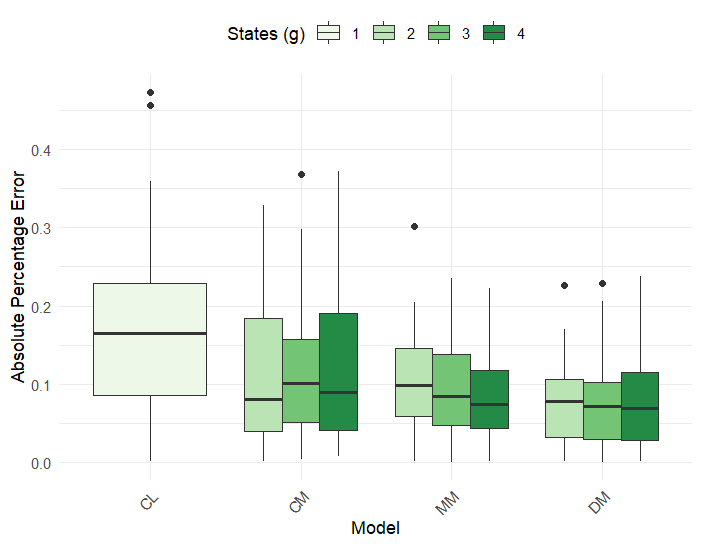}
    \caption{Boxplots of Absolute Percentage Errors for IBNR Claim Count Estimates by Model and Number of States}
    \label{fig:boxplot}
\end{figure}

To evaluate the performance of each model, we calculated the absolute percentage error of the IBNR claim count point estimates at each of the 36 valuation points. The results are summarized in boxplots that display the distribution of absolute percentage errors for each model across the valuation period, as well as in a table that reports the mean, median, and standard deviation of these errors (see Figure \ref{fig:boxplot} and Table \ref{table:errors}). We considered models with 2, 3, and 4 states for our HMMs. While more than 4 states could offer additional granularity, it may be excessive for our relatively short monthly time series.

\begin{table}[t]
\centering
\begin{tabular}{lcccc}
\hline
\multicolumn{1}{|l|}{}                                  & \multicolumn{4}{c|}{$g$ = 2}                                                                                                                        \\ \hline
\multicolumn{1}{|l|}{\textbf{Models}}                   & \multicolumn{1}{c|}{CL}     & \multicolumn{1}{c|}{CM}     & \multicolumn{1}{c|}{MM}     & \multicolumn{1}{c|}{DM}     \\ \hline
\multicolumn{1}{|l|}{\textbf{Mean Absolute \% Error}}   & \multicolumn{1}{c|}{0.1689} & \multicolumn{1}{c|}{0.1127} & \multicolumn{1}{c|}{0.1021} & \multicolumn{1}{c|}{0.0784} \\ \hline
\multicolumn{1}{|l|}{\textbf{Median Absolute \% Error}} & \multicolumn{1}{c|}{0.1648} & \multicolumn{1}{c|}{0.0807} & \multicolumn{1}{c|}{0.0978} & \multicolumn{1}{c|}{0.0785} \\ \hline
\multicolumn{1}{|l|}{\textbf{SD of Absolute \% Error}}  & \multicolumn{1}{c|}{0.1171} & \multicolumn{1}{c|}{0.0901} & \multicolumn{1}{c|}{0.0619} & \multicolumn{1}{c|}{0.0537}  \\ \hline
                                                        & \multicolumn{1}{l}{}        & \multicolumn{1}{l}{}        & \multicolumn{1}{l}{}        & \multicolumn{1}{l}{}       \\ \hline
\multicolumn{1}{|l|}{}                                  & \multicolumn{4}{c|}{$g$ = 3}                                                                                                                        \\ \hline
\multicolumn{1}{|l|}{\textbf{Models}}                   & \multicolumn{1}{c|}{CL}     & \multicolumn{1}{c|}{CM}     & \multicolumn{1}{c|}{MM}     & \multicolumn{1}{c|}{DM}     \\ \hline
\multicolumn{1}{|l|}{\textbf{Mean Absolute \% Error}}   & \multicolumn{1}{c|}{0.1689} & \multicolumn{1}{c|}{0.1121} & \multicolumn{1}{c|}{0.0931} & \multicolumn{1}{c|}{0.0774} \\ \hline
\multicolumn{1}{|l|}{\textbf{Median Absolute \% Error}} & \multicolumn{1}{c|}{0.1648} & \multicolumn{1}{c|}{0.1012} & \multicolumn{1}{c|}{0.0848} & \multicolumn{1}{c|}{0.0720} \\ \hline
\multicolumn{1}{|l|}{\textbf{SD of Absolute \% Error}}  & \multicolumn{1}{c|}{0.1171} & \multicolumn{1}{c|}{0.0875} & \multicolumn{1}{c|}{0.0631} & \multicolumn{1}{c|}{0.0630} \\ \hline
                                                        & \multicolumn{1}{l}{}        & \multicolumn{1}{l}{}        & \multicolumn{1}{l}{}        & \multicolumn{1}{l}{}       \\ \hline
\multicolumn{1}{|l|}{}                                  & \multicolumn{4}{c|}{$g$ = 4}                                                                                                                        \\ \hline
\multicolumn{1}{|l|}{\textbf{Models}}                   & \multicolumn{1}{c|}{CL}     & \multicolumn{1}{c|}{CM}     & \multicolumn{1}{c|}{MM}     & \multicolumn{1}{c|}{DM}     \\ \hline
\multicolumn{1}{|l|}{\textbf{Mean Absolute \% Error}}   & \multicolumn{1}{c|}{0.1689} & \multicolumn{1}{c|}{0.1166} & \multicolumn{1}{c|}{0.0875} & \multicolumn{1}{c|}{0.0760} \\ \hline
\multicolumn{1}{|l|}{\textbf{Median Absolute \% Error}} & \multicolumn{1}{c|}{0.1648} & \multicolumn{1}{c|}{0.0887} & \multicolumn{1}{c|}{0.0747} & \multicolumn{1}{c|}{0.0685} \\ \hline
\multicolumn{1}{|l|}{\textbf{SD of Absolute \% Error}}  & \multicolumn{1}{c|}{0.1171} & \multicolumn{1}{c|}{0.0914} & \multicolumn{1}{c|}{0.0597} & \multicolumn{1}{c|}{0.0560} \\ \hline
\end{tabular}
\caption{Summary of Absolute Percentage Errors for IBNR Claim Count Estimates Across Models}
\label{table:errors}
\end{table}

\subsubsection*{Performance Comparison of the Chain Ladder Method and HMM-Based Models}

The classical Chain Ladder (CL) method performs worse than our proposed HMM-based models. The CL method shows a mean and median absolute percentage error of around 16\%, with high variability in the error, as indicated by both the boxplot and the standard deviation values presented in Table \ref{table:errors}. This highlights the limitations of the CL method compared to more advanced approaches.

\subsubsection*{Performance of the Proposed Models}

Our proposed models—the Continuous-time Model (CM), the Multinomial Model (MM), and the Dirichlet-Multinomial Model (DM)—perform much better than the CL method, demonstrating their effectiveness in estimating IBNR claim counts. The discrete-time models (MM and DM) generally outperform the continuous-time model (CM), as evidenced by their lower median and mean absolute percentage errors and lower variability (see Figure \ref{fig:boxplot}) across all values of $g$ (2, 3, and 4). The CM model shows a median absolute error close to that of the MM and DM models, but the mean absolute error is significantly larger due to its high variability, as reflected in the boxplot.

In general, the CM model tends to produce good estimates of IBNR claim counts when the integral
\[
\int_{d_{l-1}}^{d_l} F_{U\mid t,\boldsymbol{x}_i}(\tau-t) \, dt,
\]
representing the probability that a claim occurring within the period $(d_{l-1}, d_l]$ is reported before the valuation date $\tau$, aligns closely with the actual percentage of such claims. However, when this integral deviates significantly from the actual percentage, the CM model's estimates deteriorate. For example, at the end of January 2016, the average value of the integral 
\[
\int_{d_{T-1}}^{d_T} F_{U\mid t,\boldsymbol{x}_i}(\tau-t) \, dt,
\]
where $d_T =\tau$ (which represents the probability that a claim that occurred in January 2016 was reported before the end of the month), across policyholders was 0.81095, while the actual percentage was 0.7465753 (i.e., 75\% of such claims were reported by the end of the month)—a difference that diverged from the usual pattern. This discrepancy, which was not captured by our model for reporting delay despite using all available data, significantly impacted the IBNR claim count estimate; the underestimated value of 
\[
1-\int_{d_{T-1}}^{d_T} F_{U\mid t,\boldsymbol{x}_i}(\tau-t) \, dt
\]
led to an underestimated IBNR claim count.

\subsubsection*{Effect of Increasing States on Model Performance}

As the number of states $g$ increases, the performance of the discrete-time models (MM and DM) improves, suggesting that these models benefit from more states in capturing the underlying dynamics of the data. However, this trend does not hold for the continuous-time model (CM). The CM model's performance remains inconsistent due to its sensitivity to the fitting of reporting delays and the two-step maximization of the likelihood process, as explained in the previous analysis. Therefore, while adding states improves the accuracy of discrete-time models, it does not necessarily translate into better performance for the continuous-time model.

\subsubsection*{Analysis of Confidence Intervals for DM and MM Models}

\begin{figure}[h]
    \centering
    \includegraphics[width=\textwidth]{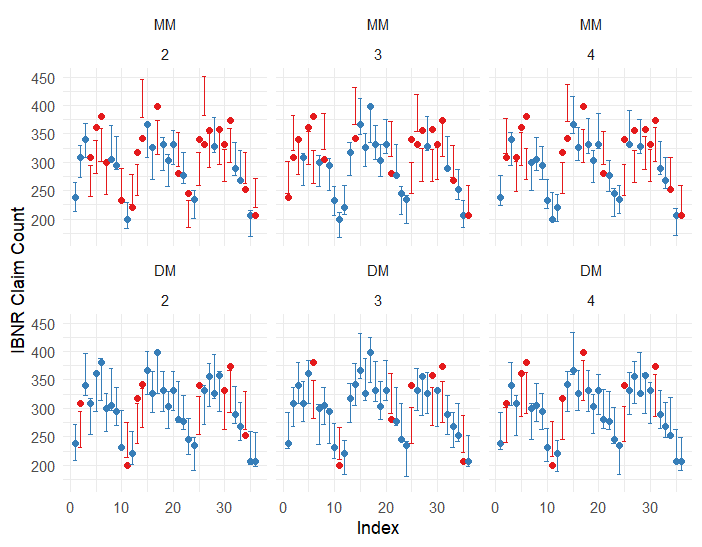}
    \caption{Error bar plots showing the 95\% confidence intervals for simulated IBNR claim count estimates using the DM and MM models with \( g = 2, 3, 4 \). The red points indicate when the actual IBNR claim count falls outside the interval, and the blue points indicate when it falls within the interval.}
    \label{fig:errorbars}
\end{figure}

The error bar plots in Figure \ref{fig:errorbars} illustrate the 95\% confidence intervals for the simulated IBNR claim count estimates using the DM and MM models with \( g = 2, 3, 4 \), while the points represent the actual IBNR claim counts. The red points indicate instances where the actual IBNR claim count falls outside the interval, and the blue points indicate instances where it falls within the interval. It is evident that the DM model's confidence intervals are generally wider and more frequently contain the actual IBNR claim counts compared to the MM model. Specifically, for the DM model, the confidence intervals include the actual IBNR claim count in 28, 29, and 28 out of the 36 valuation dates for \( g = 2, 3, 4 \), respectively. In contrast, the MM model's intervals contain the actual IBNR claim counts only 17, 20, and 21 times for \( g = 2, 3, 4 \), respectively. The increased width of the confidence intervals in the DM model reflects the introduction of variability in the reporting probability vector through the Dirichlet assumption. This added variability leads to a more appropriate distribution of the IBNR claim counts, capturing the inherent uncertainty and providing a more realistic estimate. This highlights the importance of considering such variability for better risk management, as underestimating or overestimating reserves can have significant financial implications.

\section{Conclusion} \label{sec:conc}

In this paper, we proposed a novel micro-level Cox model designed to enhance the estimation of IBNR claim counts in P\&C insurance. Our model incorporates a HMM to capture the temporal dependencies in the claim arrival process, allowing for the integration of policyholder-level data and environmental factors. This framework supports analysis at varying levels of granularity, providing flexibility for different applications. We initially presented the model in a continuous-time framework, then extended it to a discrete-time framework to enable simultaneous modeling of claim occurrence and reporting delays. Additionally, we introduced a Dirichlet distribution assumption for the reporting delay probabilities, addressing non-systematic or unexplainable variations in delay structures.

Our empirical analysis demonstrated that the discrete-time models, particularly the one incorporating the Dirichlet assumption, outperformed the continuous-time model, providing more accurate estimates of the IBNR claim count, with both frameworks outperforming the classical Chain Ladder method. This finding highlights the importance of jointly modeling claim occurrence and reporting behavior, as well as the value of accounting for randomness in reporting delays.

For future research, we plan to conduct a deeper analysis of the applicability of continuous-time models in the literature. Given the superior performance of our discrete-time approach, it is essential to reassess the contexts in which continuous-time models are most appropriate. Additionally, the inclusion of the Dirichlet assumption in our model suggests a promising direction for the application of Bayesian methods in IBNR reserving. By transitioning to a Bayesian framework, we aim to further improve the robustness and flexibility of our reserve estimates.

\textbf{Competing interests:} The authors declare none
\bibliographystyle{apalike}
\bibliography{references}

\appendix
\section{EM Algorithm for the Continuous-time Model} \label{appendixA}

Recall that the likelihood of the observed data for the continuous-time model, $\mathcal{L}^{(1)}$, is given by:
\begin{equation*} 
\mathcal{L}^{(1)}(\boldsymbol{\Phi}_1|\mathcal{O}_C) \propto P\left(\boldsymbol{N}_1^{(\text{r},1:T)}=\boldsymbol{n}_1^{(\text{r},1:T)},\dots,\boldsymbol{N}_m^{(\text{r},1:T)}=\boldsymbol{n}_m^{(\text{r},1:T)}\right) \times \prod_{i=1}^m \prod_{t=1}^T \prod_{j=1}^{n_{i,t}^\text{r}} \frac{f_{U|t_{itj},\boldsymbol{x}_i}(u_{itj})}{F_{U|t_{itj},\boldsymbol{x}_i}(\tau-t_{itj})},
\end{equation*} 
where $\boldsymbol{N}_{i}^{(r,1:T)}  = (N_{i,1}^r,\dots,N_{i,T}^r)$ denote the discretely observed reported claim process from period 1 to $T$ for policy $i$, and $\boldsymbol{n}_{i}^{(r,1:T)}=(n_{i,1}^r,\dots,n_{i,T}^r)$ is its realization. We maximize the likelihood by employing the two-step maximization widely used in the micro-level reserving literature. To this end, we describe the EM algorithm needed to fit the `Claim Frequency` component assuming a known distribution for the reporting delay. Recall that the likelihood such component is given by
$$P\left(\boldsymbol{N}_{1}^{(r,1:T)}=\boldsymbol{n}_{1}^{(r,1:T)},\dots,\boldsymbol{N}_{m}^{(r,1:T)}=\boldsymbol{n}_{m}^{(r,1:T)}\right)=\boldsymbol{\pi}_1\boldsymbol{P}_1(\boldsymbol{n}_1^r)\boldsymbol{\Gamma}\boldsymbol{P}_2(\boldsymbol{n}_2^r)\dots\boldsymbol{\Gamma}\boldsymbol{P}_T(\boldsymbol{n}_T^r)\mathbf{1}^T=:L_T,$$
where 
$$\boldsymbol{n}_{t}^\text{r}  =(n_{1,t}^\text{r},\dots,n_{m,t}^\text{r}), $$  $$\boldsymbol{P}_t(\boldsymbol{n}_t^r) = \text{diag}\left\{\prod_{i=1}^m \boldsymbol{P}(N_{i,t}^r=n_{i,t}^r|C_t=1),\dots,\prod_{i=1}^m \boldsymbol{P}(N_{i,t}^r=n_{i,t}^r|C_t=g)\right\},$$
and $\mathbf{1}$ is a row vector of 1s. 

\subsubsection*{Complete Log-Likelihood}

The complete data is given by the observed data (the discretely observed reported claims) and the unobserved data (the states of the HMM). The log-likelihood of the complete data is given by
\begin{align*}
     &l_{\text{C}}^{(1)}(\boldsymbol{\Phi}_1|\boldsymbol{N}_{i}^{(r,1:T)}=\boldsymbol{n}_{i}^{(r,1:T)}, C_t=c_t,i=1,\dots,m, t = 1,\dots,T) \\
     &=\log\left(\pi_{1,c_1}\prod_{t=2}^T\gamma_{c_{t-1},c_t}\prod_{t=1}^T \prod_{i=1}^m \mathbb{P}_{c_t}(n_{i,t}^r)\right)  =\log \pi_{1,c_1}+ \sum_{t=2}^T\log \gamma_{c_{t-1},c_t} + \sum_{t=1}^T \sum_{i=1}^m \log \mathbb{P}_{c_t}(n_{i,t}^r) \\
    & =  \sum_{j=1}^g u_{1j}\log \pi_{1j}  + \sum_{j=1}^g\sum_{k=1}^g\left(\sum_{t=2}^Tv_{tjk} \right)\log \gamma_{jk} + \sum_{j=1}^g\sum_{t=1}^T \sum_{i=1}^m u_{tj}\log \mathbb{P}_{j}(n_{i,t}^r). 
\end{align*}
where $\mathbb{P}_{j}(n_{i,t}^r) = P(N_{i,t}^r=n_{i,t}^r|C_t = j)$. 

\subsubsection*{E-step}
In the E-step, we compute the conditional expectation of the complete log-likelihood given the observed data and the current estimator $\boldsymbol{\Phi}_1^{(k-1)}$ of the model parameters $\boldsymbol{\Phi}_1$. This conditional expectation is given by:

\begin{equation} \label{Estep-eqn}
\begin{split}
        \mathbb{E}\left(l^{(1)}(\boldsymbol{\Phi}_1|\boldsymbol{n}^{(r,1:T)},\boldsymbol{c}^{(T)})|\boldsymbol{n}^{(r,1:T)},\boldsymbol{\Phi}_1^{(k-1)}\right)  & = \sum_{j=1}^g \hat{u}_{1j}^{(k)}\log \pi_{1j}  + \sum_{j=1}^g\sum_{l=1}^g\left(\sum_{t=2}^T\hat{v}_{tjl}^{(k)}\right)\log \gamma_{jl}\\& + \sum_{j=1}^g\sum_{t=1}^T \sum_{i=1}^m \hat{u}_{tj}^{(k)}\log \mathbb{P}_{j}(n_{i,t}^r),
\end{split}
\end{equation}

where $\hat{u}_{tj}^{(k)}$ and $\hat{v}_{tjl}^{(k)}$ are given by Equations \ref{eqn:uhat} and \ref{eqn:vhat}, respectively.

\subsubsection*{M-step}

The M-step in the EM algorithm involves maximizing Equation \ref{Estep-eqn} with respect to the model parameters $\boldsymbol{\Phi}_1=\{\boldsymbol{\pi}_1,\boldsymbol{\Gamma},\boldsymbol{\theta}_1,\dots,\boldsymbol{\theta}_g\}$, subject to the necessary constraints. The optimal values for $\pi_{1j}^{(k)}$ and $\gamma_{jl}^{(k)}$ at the $k$th iteration are given by Equations \ref{eqn:pi} and \ref{eqn:gamma}, respectively. Finally, we aim to obtain the regression parameter $\boldsymbol{\theta}_j^{(k)}$ that maximizes the third term of Equation \ref{Estep-eqn}, given by:  
$$\sum_{t=1}^T \sum_{i=1}^m \hat{u}_{tj}^{(k)}\log \mathbb{P}_{j}(n_{i,t};\boldsymbol{\theta}_j) = \sum_{t,i}\hat{u}_{tj}^{(k)}\log \mathbb{P}_{j}(n_{i,t};\boldsymbol{\theta}_j),$$
which is the likelihood for a weighted Poisson regression. The maximization of this likelihood is readily available in many software packages in R.

\textbf{Remark:} To expedite the maximization process, one can employ the binning technique for continuous covariates, treating them as nominal covariates. This approach significantly enhances the computational efficiency of the estimation. By creating categorical classes of policies based on binned covariates, we assume that policies within each class share similar risk characteristics. For an even faster maximization process, we can impose a restriction while binning continuous covariates, ensuring that each class has at least one claim occurrence for some $t\in{1,\dots,T}$. This enables us to express the likelihood function as a weighted exponential likelihood regression. Consequently, the maximization step becomes notably faster and more manageable, allowing us to efficiently estimate the model parameters in just a few minutes.

\subsubsection*{Convergence Criterion, Model Initialization \& Model Selection}
We use similar procedures as those for the discrete-time models.

\end{document}